\documentclass[11pt,english]{article}
\usepackage{amsmath}
\usepackage{amssymb}
\usepackage{fontspec}
\usepackage{unicode-math}
\usepackage{color}
\usepackage{array}
\usepackage{booktabs}
\usepackage[letterpaper]{geometry}
\usepackage{microtype}
\usepackage[pdfusetitle,
 bookmarks=true,bookmarksnumbered=true,bookmarksopen=true,bookmarksopenlevel=3,
 breaklinks=true,pdfborder={0 0 0},pdfborderstyle={},backref=false,colorlinks=true]
 {hyperref}
\hypersetup{
 linktoc=all}

\makeatletter

\numberwithin{equation}{section}
\numberwithin{figure}{section}
\numberwithin{table}{section}

\@ifundefined{date}{}{\date{}}
\usepackage[margin=1cm]{caption}
\usepackage{tocloft}
\cftsetindents{subsubsection}{4em}{4em}
\DeclareMathOperator{\e}{e}
\DeclareMathOperator{\ii}{i}
\DeclareMathOperator{\tr}{tr}
\usepackage{fontawesome5}
\DeclareMicrotypeAlias{texgyrepagella-regular.otf}{Palatino}
\AtBeginDocument{\setcounter{biburlnumpenalty}{9000}\setcounter{biburlucpenalty}{9000}\setcounter{biburllcpenalty}{9000}}
\AtBeginDocument{\AtBeginBibliography{\setlength{\emergencystretch}{3em}}}

\makeatother

\usepackage{polyglossia}
\setdefaultlanguage[variant=american]{english}
\usepackage[style=numeric-comp,sorting=none,giveninits=false,doi=true,eprint=true,backref=true]{biblatex}

\begin{document}
\global\long\def\BBZ{\mathbb{Z}}%
\global\long\def\HH{\mathcal{H}}%
\global\long\def\dn{\downarrow}%
\global\long\def\hf{\frac{1}{2}}%
\global\long\def\fa{\forall}%
\global\long\def\pld{\dot{+}}%
\global\long\def\mnd{\dot{-}}%

\title{Time Travel Paradoxes and Entangled Timelines}
\author{\textbf{Barak Shoshany}{\small}\\
{\small{} \faIcon{envelope} \href{mailto:bshoshany@brocku.ca}{bshoshany@brocku.ca}$\qquad$\faIcon{orcid}
\href{https://orcid.org/0000-0003-2222-127X}{0000-0003-2222-127X}$\qquad$\faIcon{globe}
\href{https://baraksh.com/}{https:/\kern-0.3em/baraksh.com/}}\\
{\small\faIcon{university} \href{https://brocku.ca/}{Brock University}}\\
\\
\textbf{Zipora Stober}\\
{\small{} \faIcon{envelope} \href{mailto:zipora@ad.unc.edu}{zipora@ad.unc.edu}$\qquad$\faIcon{orcid}
\href{https://orcid.org/0000-0001-5552-6593}{0000-0001-5552-6593}}\\
{\small\faIcon{university} \href{https://brocku.ca/}{Brock University}
}{\small\emph{and}}{\small{} \href{https://www.unc.edu/}{The University of North Carolina at Chapel Hill}}}
\maketitle
\begin{abstract}
For time travel to be consistent with the known laws of physics, the resulting paradoxes must be resolved. It has been suggested that the notion of parallel timelines may provide a resolution, by allowing different events to happen in different timelines. However, so far, a concrete mechanism by which parallel timelines can be created has never been satisfactorily formulated in a way that makes the relationships between the timelines clear. In this paper, we propose such a mechanism based on the Everett (many-worlds) interpretation of quantum mechanics. The timelines in our model are variants of the ``worlds'' of the Everett interpretation that can connect to each other via the time machine. They are created by the entanglement of the time machine with the environment and the resulting decoherence; therefore, we call them ``entangled timelines'' or E-CTCs. As the entanglement gradually spreads out to additional systems, the timelines spread out as well, providing a local and well-defined mechanism for branching timelines. The E-CTC model is similar to Deutsch's familiar D-CTC model but differs from it mainly by imposing a different consistency condition, based on pure states, which allows the information about the entanglement to be preserved. These differences allow us to create a clearer practical definition of the timelines and avoid some of the surprising consequences of the D-CTC model's non-linear evolution.
\end{abstract}

\tableofcontents{}

\section{Introduction}

\subsection{Time Travel and Its Paradoxes}

General relativity \cite{Hawking1973,Wald1984,Carroll2004} is our most successful and precise theory of spacetime and causality. This theory admits solutions to the Einstein equations \cite{Godel1949} containing \textbf{closed timelike curves (CTCs)}, which can hypothetically be used to travel to the past and violate causality \cite{Shoshany2019,Lobo2017,Krasnikov2018}. Such curves may in principle arise in spacetime geometries that allow effective faster-than-light travel, such as wormholes \cite{Shoshany2023,Shoshany2025} and warp drives \cite{Shoshany2024,Alcubierre1994}. It is currently unknown whether such spacetime geometries are permissible in our universe, as they appear to require exotic matter with negative energy density \cite{Shoshany2019,Lobo2017,Krasnikov2018}, and it is not known whether such matter can be created in sufficient quantities.

The concept of \textbf{closed causal curves (CCCs)}, where ``causal'' means non-spacelike, provides a notion of time travel more general than CTCs. Although a closed lightlike curve cannot be used by an observer to physically travel to the past, it can still be used to send information to the past and create paradoxes. However, in the literature, CTCs have become synonymous with time travel, so we stick to that terminology here as well.

If CTCs can be created, and time travel is possible, this would lead to time travel paradoxes, which must be resolved \cite{Shoshany2020,Shoshany2023,Visser1996,Krasnikov1997,Krasnikov2002,Wasserman2018}. Two main types of time travel paradoxes frequently appear in the theoretical physics literature. The first type is \textbf{consistency paradoxes}, where time travel to the past creates a chain of events that ends up preventing the journey through time from happening in the first place.

For example, if Alice puts a bomb inside the time machine and sends it a few minutes to the past, then when the bomb arrives, it will kill Alice, or even destroy the time machine itself; but in that case, Alice will not be able to send the bomb through the time machine after all. This chain of events is inconsistent, and hence paradoxical.

The second type of time travel paradoxes is \textbf{bootstrap paradoxes}, where an event causes itself---or more precisely, the chain of events is a closed loop, with no cause outside the loop \cite{Lossev1992}. For example, consider the scenario where Bob is working on a new book but struggling with writer's block. He builds a time machine, opens it, and finds the finished book inside.

Bob publishes the book and becomes a best-selling author. A few years later, he sends the book to his past self using the time machine. In this case, everything is perfectly consistent, so we do not have a consistency paradox; and yet, the entire chain of events has no external cause, and the book appears to have been created out of nothing.

One could argue that bootstrap paradoxes are of no real concern, as the lack of external cause means there is no reason for the closed loop of events to exist in the first place \cite{Krasnikov2018}. However, in \cite{Shoshany2023} we gave some motivation for taking bootstrap paradoxes more seriously, in the form of ``the password paradox'', where the mere requirement of self-consistency might by itself serve as the external cause giving rise to the closed loop.

In the case of consistency paradoxes, on the other hand, it is generally agreed upon that they present a significant difficulty. Such paradoxes are usually depicted in science fiction as actions that must be avoided by the characters; but in reality, paradoxes in physics usually indicate an inconsistency in the theory itself. Hence, in order to formulate a consistent fundamental theory of physics, it is crucial to either:
\begin{enumerate}
\item Prove beyond any doubt that time travel is impossible;
\item Provide a proper resolution to the paradoxes it creates.
\end{enumerate}

Option (1) is the statement of the \textbf{Hawking chronology protection conjecture} \cite{Gott1991,Hawking1992,Visser1993,Grant1993,Visser2003,Earman2009}. Attempts to prove this conjecture generally utilize quantum field theory on curved spacetime to show that some physical quantities are ill-defined or diverge in the presence of time machines due to quantum effects \cite{Kim1991,Kay1997}. However, so far, such proofs only appear to work in some cases and not in others \cite{Krasnikov1996,Krasnikov1999}.

Furthermore, there is no reason to believe that current techniques using quantum field theory on curved spacetime as an approximation to quantum gravity are valid in the presence of CTCs, and thus any proofs relying on issues that arise in such cases could merely be exposing issues with these techniques, rather than with time travel itself. Hence, a conclusive proof of the Hawking conjecture may not be possible without a consistent and experimentally verified theory of quantum gravity.

At the moment, Hawking's conjecture remains unproven, and time travel remains a possibility. Even if the conjecture is eventually proven, we believe there is something to be learned about time and causality in general by studying time travel paradoxes even if time travel itself is impossible. Hence, in this paper, we will focus on option (2).

\subsection{Proposed Resolutions to the Paradoxes}

One proposed way to resolve time travel paradoxes, known as the \textbf{Novikov self-consistency conjecture} \cite{Friedman1990,Novikov1992}, suggests that one can simply never make any changes to the past. Any attempts to change the past will necessarily fail, or even bring about the very future they tried to prevent. If the past cannot be changed, then there is also no possibility of paradoxes.

In more technical terms, Novikov's conjecture proposes that local solutions to the equations of motion must also be globally self-consistent. In other words, any attempt to create an inconsistency, despite seemingly being possible according to all \textbf{local} laws of physics, will not work because it will violate \textbf{global} consistency. This means that the probability of any event that would cause a paradox must be exactly zero, and only initial conditions leading to self-consistent evolutions are allowed.

In the consistency paradox described above, where Alice tries to kill her past self by sending a bomb back in time, the self-consistency conjecture implies that this goal cannot possibly be achieved. Perhaps the bomb fails to explode, or perhaps it does explode, but past-Alice survives. The bootstrap paradox, however, cannot be resolved by the self-consistency conjecture, as it is fully compatible with it---there are no inconsistencies.

It has been shown that in some cases, this conjecture can indeed be used to resolve paradoxes \cite{Friedman1990,Echeverria1991,Novikov1992,Carlini1995}. Nevertheless, in three previous publications \cite{Shoshany2019,Shoshany2020,Shoshany2023}, we proposed several arguments against the validity of the self-consistency conjecture. Most importantly, we have shown that there are some concrete time travel paradox models that simply cannot be made self-consistent, as they generate an inconsistency with probability 1 for any initial condition.

Another possible resolution of time travel paradoxes discussed in the literature is \textbf{parallel timelines}, also known as \textbf{multiple histories}. In this scenario, time travel necessarily results in creating a new timeline, or equivalently, splitting one timeline into two. In Refs.~\cite{Shoshany2023,Shoshany2020}, we assumed the existence of parallel timelines and showed that they resolve the paradoxes that Novikov's conjecture cannot.

Let us consider how the consistency paradox is resolved in the parallel timelines approach. Alice is alive in timeline 0, and therefore she is able to send a bomb back in time. The bomb arrives in a separate timeline, timeline 1, where it kills Alice. However, the fact that Alice is dead in timeline 1 does not create an inconsistency, as the bomb was sent by Alice of timeline 0, who is alive and well.

The bootstrap paradox can also easily be resolved using parallel timelines. In timeline 0, Bob works hard and eventually finishes writing his book. He then sends the book back in time. The book arrives in a separate timeline, timeline 1, where Bob benefits from it without having to do the work. However, the book was clearly not created from nothing; it was created by Bob of timeline 0.

\subsection{Quantum Mechanics and Entangled Timelines}

The main issue with the parallel timelines resolution to time travel paradoxes is that a concrete and rigorous model incorporating them into both general relativity and quantum mechanics is lacking.

On the general relativity front, it has been postulated that either non-Hausdorff manifolds \cite{Hajicek1971,Sharlow1998,Muller2013,Placek2014,Luc2019,Luc2020} or spaces that are not locally Euclidean \cite{McCabe2005} may allow spacetime to ``branch'' into several physically distinct futures when time travel occurs. However, such models present considerable mathematical difficulties. Even if these difficulties are resolved, it is still unclear what physical mechanism would cause spacetime to branch in the first place, or where the space and matter making up the new branch are supposed to come from.

On the quantum mechanics front, Deutsch \cite{Deutsch1991} famously showed that certain time travel paradoxes, formulated in terms of qubits and quantum gates, can be resolved by demanding a consistency condition on the time-traveling portion of the quantum state, obtained by taking a partial trace. This is now known as \textbf{the D-CTC model}.

Deutsch suggests that the most sensible way to interpret this model is using purely unitary quantum mechanics, also known as \textbf{the Everett or ``many-worlds'' interpretation (MWI)} \cite{Everett1957,DeWitt1970,Saunders2010,Everett2012,Wallace2012,Vaidman2024}. Therefore, the D-CTC model can be seen as a quantum model of parallel timelines \cite{Deutsch1994}. However, as we will discuss in detail in Section~\ref{sec:Comparison-of-our}, the model has several significant drawbacks when interpreted in this way. In particular, the consistency condition imposed in the D-CTC model ends up obscuring important details about the timelines.

In this paper, we propose a new model for resolving time travel paradoxes using parallel timelines. Our model is also based on the MWI, but it is different from the D-CTC model, and avoids the issues outlined above. In our model, the parallel timelines are created explicitly through entanglement and decoherence involving the time machine and its environment, similarly to the ``worlds'' of the MWI. For this reason, we named this model \textbf{``entangled timelines''} or \textbf{the E-CTC model}. Here we continue the ``X-CTC'' naming convention that has become prevalent in the literature on time travel models, with another prominent example being P-CTC, which stands for ``post-selected CTC'' \cite{Lloyd2011a,Lloyd2011b}. We do not discuss the P-CTC model in this paper, as it is not a parallel-timelines model.

The entangled timelines are an \textbf{emergent} concept. Each timeline is not a separate universe but rather a separate decohered component of the overall quantum state of a single universe. The chain of events within each timeline can be followed continuously via the action of an evolution operator, and each timeline can be related to the next one via the action of a correlation operator.

Time travel paradoxes are closely related to the notion of causality, which is necessarily disrupted in the presence of CTCs. Causality is usually thought of as a classical concept, and its status in quantum mechanics is more subtle. In recent years, quantum causality and processes with no definite causal order have been studied extensively \cite{Oreshkov2012,Chiribella2013,Brukner2014,Zych2019}, including in relation to time travel \cite{Araujo2017,Baumeler2019}. In this paper we take a more conservative approach, which is standard in the literature on the D-CTC model: the CTC is part of a fixed classical spacetime, and only the systems interacting with it are treated quantum mechanically. As we will see, since the timelines in our model emerge via decoherence, each timeline is quasi-classical, and the events within it have the usual causal order. Every event still has a cause, but the cause of an object emerging from the time machine is located in a different timeline.

In this paper, we consider a simple paradox model in both the D-CTC and E-CTC frameworks. We show that the mixed states obtained in the D-CTC resolution are the reduced states of the corresponding pure entangled E-CTC state. This reduction discards information about the entanglement and therefore about the timelines themselves. We thus argue that our model is better suited for understanding how timelines naturally emerge from the MWI.

Entangled timelines are created locally when a system interacts with the time machine and decoherence suppresses interference between the resulting components of the entangled state. They then gradually spread throughout the causal future of the interaction as more systems become entangled. This provides a new and simpler way to think about the topology and geometry of spacetimes with parallel timelines. Unlike the branching spacetime picture mentioned above, where the entire spacetime branches into several physical copies of itself in order to accommodate the different timelines, the entangled timelines picture defines timelines as emergent entities via local interactions between systems within a single \textbf{non-branching} spacetime.

\section{The Paradox Model}

\subsection{\label{subsec:The-generic-paradox}The Generic Paradox}

In previous publications \cite{Shoshany2020,Shoshany2023}, we used paradox models with particular physical objects in a specific spacetime geometry, using properties of the objects, such as charge or temperature, to create a paradox. However, here we would like to instead keep things as general as possible. Therefore, we must not limit ourselves to one spacetime geometry or one particular physical system.

The generic setup for our model will be as follows. The time machine is represented by a Hilbert space $\HH_{\mathrm{CTC}}$, which encodes the possible quantum states of the time machine and which objects, if any, have passed through it. The time machine can, for example, be an identification of two regions of spacetime, with the proper precautions taken to avoid singularities, horizons, and other issues. This can be described precisely using the Morris-Thorne wormhole or other traversable wormholes with their mouths separated in time \cite{Morris1988a,Morris1988b,Visser1996}.

However, the exact implementation of the time machine is not important; all we care about is that there is a CTC somewhere. The topology and geometry of the spacetime containing this CTC, and the shape of the CTC itself, are irrelevant for our purposes. We simply assume that there is a way to move an object along a CTC back to an earlier point in time.

Aside from the time machine, we also have a Hilbert space $\HH_{\mathrm{ex}}$ external to the time machine. This Hilbert space may describe a particle, a billiard ball, a human, or any other physical system that we would like to use to create the paradox. We assume that the external system and the time machine are initially uncorrelated; that is, their joint state is a product state. However, once the two systems interact, their states can (and will) be correlated.

We represent both Hilbert spaces as qubits in the computational basis:
\begin{itemize}
\item For $\HH_{\mathrm{CTC}}$, $\left|0\right\rangle $ means the time machine is \textbf{empty} and $\left|1\right\rangle $ means the time machine is \textbf{not empty}. This qubit can be interpreted as an occupation number, where an object (particle, ball, human) is either absent or present inside the time machine.
\item For $\HH_{\mathrm{ex}}$, $\left|0\right\rangle $ means \textbf{time travel will not occur} and $\left|1\right\rangle $ means \textbf{time travel will occur}. This qubit can be interpreted as a control bit deciding whether the time machine will operate, independently of the specific physical systems involved.
\end{itemize}

We stress that these qubit states should not be interpreted as representing the states of any physical 2-state system. Instead, they should be interpreted as representing abstract \textbf{logical} (true or false) states. In the next two sections we will give more concrete examples, in which we will replace the qubit labels with more informative ones.

As in Deutsch's model \cite{Deutsch1991}, the Hilbert space $\HH_{\mathrm{CTC}}$ is assigned to the system traveling along the CTC, and keeps track of what, if anything, has emerged from the time machine, ignoring any details that are not relevant to the paradox. In particular, the object passing through the time machine does not have to be a material one. A signal carrying information to the past is sufficient to create a paradox, and can be described using the same two states.

States of the overall system will be written as tensor products in the composite Hilbert space $\HH_{\mathrm{CTC}}\otimes\HH_{\mathrm{ex}}$ and denoted by $\left|\Psi\left(t\right)\right\rangle $, where $t$ is time. The canonical initial state, at $t=0$, is
\begin{equation}
\left|\Psi\left(0\right)\right\rangle =\left|0\right\rangle \otimes\left|1\right\rangle ,\label{eq:empty-initial}
\end{equation}
where $\left|0\right\rangle \in\HH_{\mathrm{CTC}}$ is the state of the time machine when it is empty---that is, no one has traveled through it yet---and $\left|1\right\rangle \in\HH_{\mathrm{ex}}$ is the initial state of the external system, indicating that time travel will occur by default, unless some interaction prevents it from happening.

At $t=1$, the system is still in the same state, since nothing happened:
\[
\left|\Psi\left(1\right)\right\rangle =\left|0\right\rangle \otimes\left|1\right\rangle .
\]
Since $\HH_{\mathrm{ex}}$ is still in the state $\left|1\right\rangle $ at $t=1$, time travel will indeed occur. Physically, this may mean that the actual object represented by $\HH_{\mathrm{ex}}$ (particle, ball, human, etc.) goes into the time machine, or perhaps a different object travels in time as a result of $\HH_{\mathrm{ex}}$'s state (e.g., if Alice is still alive, she can send a bomb through the time machine, as will be the case in the macroscopic example below).

Whatever the case may be, time travel has occurred, so the time machine will not be empty at $t=0$. Therefore, the state at $t=0$ after time travel has occurred will be
\[
\left|\Psi\left(0\right)\right\rangle =\left|1\right\rangle \otimes\left|1\right\rangle .
\]
The state of $\HH_{\mathrm{CTC}}$ is now $\left|1\right\rangle $, indicating that the time machine is not empty. However, $\HH_{\mathrm{ex}}$ \textbf{always} starts in the same state $\left|1\right\rangle $, indicating that time travel \textbf{may} happen unless something prevents it from happening before $t=1$.

To create a paradox, we demand that whenever the time machine is not empty, that is, the state of $\HH_{\mathrm{CTC}}$ is $\left|1\right\rangle $, time travel is prevented. This could be, for example, a particle annihilating its past self, or a bomb killing the human who was supposed to enter the time machine. In any case, the state of the system at $t=1$ will now be
\[
\left|\Psi\left(1\right)\right\rangle =\left|1\right\rangle \otimes\left|0\right\rangle ,
\]
where $\left|0\right\rangle \in\HH_{\mathrm{ex}}$ represents the state where time travel has been prevented from happening; note that we will formulate the time evolution from $t=0$ to $t=1$ more precisely in section~\ref{subsec:Resolution-using-quantum}, when we discuss the paradox as a quantum system. Finally, since time travel is not happening, the time machine must be empty at $t=0$, and we get back to \eqref{eq:empty-initial}.

The entire inconsistent chain of events is summarized in Table \ref{tab:inconsistent-single-timeline}, where $\downarrow$ represents normal time evolution to the future and $\circlearrowright$ means that the state of the system at $t=1$ correlates with the state at $t=0$ written below it (for example, if at $t=1$ time travel happens, then the corresponding state at $t=0$ is one where the time machine is not empty). At the end of this chain of events, we get back to the original state at $t=0$, so the chain is \textbf{cyclic}.

\begin{table}[tbph]
\centering
\caption{The inconsistent cyclic sequence of states in the single-timeline model.}
\label{tab:inconsistent-single-timeline}
\begin{tabular}{cc}
\toprule
\textbf{Time} & \textbf{State/Event}\\
\midrule
$t=0$ & $\left|0\right\rangle \otimes\left|1\right\rangle $\\
$\dn$ & \textbf{Nothing happens}\\
$t=1$ & $\left|0\right\rangle \otimes\left|1\right\rangle $\\
$\circlearrowright$ & \textbf{Time travel occurs}\\
$t=0$ & $\left|1\right\rangle \otimes\left|1\right\rangle $\\
$\dn$ & \textbf{Time travel is prevented}\\
$t=1$ & $\left|1\right\rangle \otimes\left|0\right\rangle $\\
$\circlearrowright$ & \textbf{Time travel does not occur}\\
$t=0$ & $\left|0\right\rangle \otimes\left|1\right\rangle $\\
\bottomrule
\end{tabular}
\end{table}

If we treat this chain of events \textbf{classically}, there is clearly a consistency paradox here. The paradox comes from the fact that if time travel happens, then it prevents itself from happening. We can see the inconsistency by noting that the time machine must be in both states $\left|0\right\rangle $ and $\left|1\right\rangle $ at $t=0$, and the external system must be in both states $\left|0\right\rangle $ and $\left|1\right\rangle $ at $t=1$.

Furthermore, attempting to invoke Novikov's self-consistency conjecture will not help us, as there is no way to make this evolution consistent, as was done, e.g., by correlating the trajectories of the billiard balls in \cite{Friedman1990}. The specific physical details of the system do not matter here; \textbf{by design}, the time machine being non-empty automatically leads to time travel not happening. Therefore, this is \textbf{a true consistency paradox}: a system which simply has no consistent classical evolution.

The generic paradox presented here was created with the explicit goal of being as simple as possible. It does not describe a concrete physical system but rather the most general properties that a system which includes a time machine can have. We define the qubit states $\left|0\right\rangle $ and $\left|1\right\rangle $ in each Hilbert space based on generic conditions (time machine is empty or not, time travel happens or not), ignoring anything else that may be going on within each system but has no direct effect on the paradox itself.

We can do this for the same reason that a qubit can be encoded in the spin of an electron, with $\left|0\right\rangle $ corresponding to spin up along the $z$ direction and $\left|1\right\rangle $ corresponding to spin down; the electron has other degrees of freedom, such as position and momentum, but these are not relevant to the qubit encoding, and therefore the 2-dimensional spin Hilbert space spanned by the states $\left|0\right\rangle $ and $\left|1\right\rangle $ is sufficient.

To describe a concrete physical system using our generic paradox, the qubit states $\left|0\right\rangle $ and $\left|1\right\rangle $ in each Hilbert space can be mapped to any microscopic or macroscopic system we like. Moreover, once we resolve the paradox for the qubit states, the resolution can be straightforwardly mapped to the concrete system. In the next two sections we will present two such systems, one macroscopic and the other microscopic.

\subsection{\label{subsec:A-macroscopic-example}A Macroscopic Example: Alice and the Bomb}

We begin with a macroscopic example, since it can be understood more intuitively. However, dealing with macroscopic systems leads to many subtleties and complications, which we will discuss in detail in Section~\ref{sec:Many-worlds-and}. We are not claiming that the model we will present now is perfectly rigorous, but we still present it here simply because it is conceptually interesting and illuminating.

In the macroscopic model, Alice, a time travel physicist, has a time machine in her possession. At $t=1$, Alice puts a bomb inside the time machine and sends it back to $t=0$. The bomb is set up so that it triggers as soon as the time machine's door opens. When past-Alice opens the door, the bomb explodes, killing her. Since Alice is now dead, she cannot send a bomb back in time at $t=1$, but that means she will be alive, so she can send a bomb. This is obviously a consistency paradox. Note that in formulating this example, we decided to have Alice send a bomb instead of traveling to the past in person with plans to kill her past self, in order to avoid any philosophical issues of free will.

This example can be formulated in terms of the generic paradox by simply replacing the labels of the qubits with more descriptive ones. For $\HH_{\mathrm{CTC}}$, we define:
\begin{itemize}
\item $\left|0\right\rangle \equiv\left|\textrm{empty}\right\rangle $, the state where the time machine is empty and there is no bomb inside.
\item $\left|1\right\rangle \equiv\left|\textrm{bomb}\right\rangle $, the state where the time machine has a bomb inside.
\end{itemize}

For $\HH_{\mathrm{ex}}$, we define:
\begin{itemize}
\item $\left|0\right\rangle \equiv\left|\textrm{dead}\right\rangle $, the state where Alice was killed by the bomb.
\item $\left|1\right\rangle \equiv\left|\textrm{alive}\right\rangle $, the state where Alice is still alive.
\end{itemize}

The chain of events is given by replacing the labels in the generic table of Section~\ref{subsec:The-generic-paradox}.

\subsection{\label{subsec:A-microscopic-example}A Microscopic Example: Particle Annihilation}

As another illustration of the paradox, let us now give a microscopic example. This has the benefit of being immune to the potential issues that may arise from dealing with macroscopic systems.

We consider a species of subatomic particles with the property that whenever two of them collide, they annihilate. In this simplified example, we are not concerned with the exact nature of the annihilation or the end results; it may be, for example, that two photons are emitted in this annihilation, carrying away the initial energy and momentum, but this does not affect the paradox.

A particle is approaching the time machine. At $t=1$, it enters the time machine and comes out at $t=0$. It then collides and annihilates with its past self, preventing it from entering the time machine at $t=1$. Since the particle does not enter the time machine, the annihilation never happens. Hence, we get a consistency paradox.

For $\HH_{\mathrm{CTC}}$, we define:
\begin{itemize}
\item $\left|0\right\rangle \equiv\left|\textrm{empty}\right\rangle $, the state where no particle came through the time machine.
\item $\left|1\right\rangle \equiv\left|\textrm{particle}\right\rangle $, the state where a particle came through the time machine.
\end{itemize}

For $\HH_{\mathrm{ex}}$, we define:
\begin{itemize}
\item $\left|0\right\rangle \equiv\left|\textrm{annihilated}\right\rangle $, the state where the past particle was annihilated by colliding with the future particle.
\item $\left|1\right\rangle \equiv\left|\textrm{not annihilated}\right\rangle $, the state where the past particle was not annihilated, since no particle came through the time machine.
\end{itemize}

The chain of events is once again given by the generic table of Section~\ref{subsec:The-generic-paradox}, with the labels replaced. Note that at the second $t=1$ state (after time travel is prevented), the fact that the state of $\HH_{\mathrm{CTC}}$ is $\left|1\right\rangle \equiv\left|\textrm{particle}\right\rangle $ does \textbf{not} mean that the particle itself still exists at $t=1$; it in fact does not exist, as it was annihilated. Remember, $\HH_{\mathrm{CTC}}$ is the state of the time machine itself, and $\left|\textrm{particle}\right\rangle $ simply means that a particle has traveled through it, making no guarantees as to whether that particle still exists after it exits the time machine.

\subsection{Resolution Using Classical Parallel Timelines}

As shown in \cite{Shoshany2020,Shoshany2023}, parallel timelines can be used to resolve time travel paradoxes, even with purely classical evolution. Let us see how this would work with our generic model and the specific examples presented above.

We denote the timeline index by $h$ (for ``history''). By convention, we always take $h=0$ to represent the timeline where the time machine is empty, that is, the timeline in which time travel has not happened ``yet''. The resulting chain of events is summarized in Table \ref{tab:classical-parallel-timelines}.

\begin{table}[tbph]
\centering
\caption{The consistent cyclic sequence of states across two classical parallel timelines.}
\label{tab:classical-parallel-timelines}
\begin{tabular}{ccc}
\toprule
\textbf{Time} & \textbf{Timeline} \boldmath{$h=0$} & \textbf{Timeline} \boldmath{$h=1$}\\
\midrule
$t=0$ & $\left|0\right\rangle \otimes\left|1\right\rangle $ & $\left|1\right\rangle \otimes\left|1\right\rangle $\\
$\dn$ & \textbf{Nothing happens} & \textbf{Time travel is prevented}\\
$t=1$ & $\left|0\right\rangle \otimes\left|1\right\rangle $ & $\left|1\right\rangle \otimes\left|0\right\rangle $\\
$\circlearrowright$ & \textbf{Time travel occurs} & \textbf{Time travel does not occur}\\
\bottomrule
\end{tabular}
\end{table}

Notice that timeline $h=0$ at $t=1$, where time travel occurs, connects to timeline $h=1$ at $t=0$, where the time machine is not empty. In turn, timeline $h=1$ at $t=1$, where time travel does not occur, connects back to the original timeline $h=0$ at $t=0$, where the time machine is empty, so the chain of events is still \textbf{cyclic}, as in the single-timeline scenario above. However, now we have no inconsistencies. Time travel does not have to both happen and not happen; it can simply happen in timeline $h=0$ and not happen in timeline $h=1$.

The corresponding tables for the macroscopic and microscopic examples are obtained by substituting the labels defined in Sections \ref{subsec:A-macroscopic-example} and \ref{subsec:A-microscopic-example}. For example, in the macroscopic case, Alice of timeline $h=0$ is alive and sends a bomb back in time, while the bomb kills Alice of timeline $h=1$, who therefore cannot send a bomb.

This clearly resolves the paradox, but what is the mechanism that allows these parallel timelines to exist in the first place? In the rest of this paper, we will show that quantum mechanics, suitably interpreted, naturally provides such a mechanism, and we will discover that this has some interesting consequences that do not exist (or at least, are not obvious) in the naive classical timelines picture.

It is worth emphasizing why a quantum treatment is required here, rather than a purely classical analysis of the consistency condition. A classical analysis, in the spirit of Novikov's conjecture, can only resolve paradoxes for which a self-consistent solution exists, and as we have seen, our generic paradox has no such solution. Classical parallel timelines do resolve the paradox, but they have to be postulated by hand, along with the rules connecting them. In contrast, as we will show below, in quantum mechanics a consistent solution exists for any initial state of the external system, since superposition allows mutually exclusive chains of events to coexist. Moreover, the existence of the timelines does not have to be postulated; they emerge from entanglement and decoherence, with no need to modify spacetime itself.

\subsection{\label{subsec:Resolution-using-quantum}Quantum Superposition and Unitary Evolution}

In order to find a natural mechanism for parallel timelines, let us now consider quantum mechanics and attempt to invoke the concept of \textbf{superposition}. By studying the chain of events for the generic paradox, we see that we should allow:
\begin{itemize}
\item Two different states at $t=0$: $\left|0\right\rangle \otimes\left|1\right\rangle $ and $\left|1\right\rangle \otimes\left|1\right\rangle $.
\item Two different states at $t=1$: $\left|0\right\rangle \otimes\left|1\right\rangle $ and $\left|1\right\rangle \otimes\left|0\right\rangle $.
\end{itemize}

$\HH_{\mathrm{ex}}$ will always start in the initial condition $\left|1\right\rangle $, by construction. Therefore, at $t=0$, we do not expect $\HH_{\mathrm{ex}}$ to be in a superposition. However, $\HH_{\mathrm{CTC}}$ can certainly be in a superposition of $\left|0\right\rangle $ and $\left|1\right\rangle $:
\[
\left|\Psi\left(0\right)\right\rangle =\left(\alpha\left|0\right\rangle +\beta\left|1\right\rangle \right)\otimes\left|1\right\rangle ,
\]
where $\alpha$ and $\beta$ are arbitrary complex amplitudes such that $\left|\alpha\right|^{2}+\left|\beta\right|^{2}=1$. Expanding the tensor product, we get the \textbf{separable} (non-entangled) state
\begin{equation}
\left|\Psi\left(0\right)\right\rangle =\alpha\left|0\right\rangle \otimes\left|1\right\rangle +\beta\left|1\right\rangle \otimes\left|1\right\rangle .\label{eq:Psi-0}
\end{equation}
To describe the time evolution, let us consider a simple \textbf{unitary evolution operator} $U$ describing the interaction between the external system and the time machine. If the time machine is empty at $t=0$, that is, the state of the system is $\left|0\right\rangle \otimes\left|1\right\rangle $, then the evolution operator does not modify the state:
\[
U\left(\left|0\right\rangle \otimes\left|1\right\rangle \right)=\left|0\right\rangle \otimes\left|1\right\rangle .
\]
However, if the time machine is not empty at $t=0$, that is, the state of the system is $\left|1\right\rangle \otimes\left|1\right\rangle $, then the object leaving the time machine interacts with the external system (particles annihilate, bombs explode, etc.) such that the evolution operator changes the state to $\left|1\right\rangle \otimes\left|0\right\rangle $:
\[
U\left(\left|1\right\rangle \otimes\left|1\right\rangle \right)=\left|1\right\rangle \otimes\left|0\right\rangle .
\]
It is trivial to construct the appropriate operator---it is simply a CNOT (controlled NOT) gate \cite{Feynman1986,Nielsen2010,Wilde2017}, which flips the second qubit (the state of the external system) if and only if the control qubit (the state of the time machine) is $\left|1\right\rangle $:
\begin{equation}
U\left(\left|x\right\rangle \otimes\left|y\right\rangle \right)=\left|x\right\rangle \otimes\left|x\pld y\right\rangle ,\label{eq:unitary-U}
\end{equation}
where $x,y\in\BBZ_{2}$ and $\pld$ denotes addition modulo 2. Note that the times $t=0$ and $t=1$ are merely convenient labels for the moments at which the object emerges from the time machine and enters it, respectively. Between these two moments, the system evolves continuously according to the Schrödinger equation, and $U$ is the resulting evolution operator. The paradox only depends on the overall effect of the interaction, not on the particular Hamiltonian which generates it.

When we evolve the state $\left|\Psi\left(0\right)\right\rangle $ using this operator, linearity dictates that it acts on each term individually. Therefore, the state at $t=1$ will be
\begin{equation}
\left|\Psi\left(1\right)\right\rangle =U\left|\Psi\left(0\right)\right\rangle =\alpha\left|0\right\rangle \otimes\left|1\right\rangle +\beta\left|1\right\rangle \otimes\left|0\right\rangle .\label{eq:Psi-1}
\end{equation}
Notice that if $\alpha$ and $\beta$ are both nonzero, the state of the external system has now become \textbf{entangled} with that of the time machine.

\subsection{\label{subsec:Correlation-and-consistency}Correlation and Consistency Across Timelines}

Next, time travel occurs. To resolve the paradox, we must ensure that the joint state at $t=1$, immediately before time travel, properly correlates with the corresponding joint state at $t=0$, immediately after arrival in the past. We do this by postulating a \textbf{timeline correlation operator} $T$ as follows:
\begin{equation}
T\left(\left|x\right\rangle \otimes\left|y\right\rangle \right)=\left|y\right\rangle \otimes\left|y\mnd x\right\rangle ,\label{eq:T-op}
\end{equation}
where $\mnd$ denotes subtraction modulo 2. Note that $T$ is unitary; its inverse is
\[
T^{-1}\left(\left|x\right\rangle \otimes\left|y\right\rangle \right)=\left|x\mnd y\right\rangle \otimes\left|x\right\rangle .
\]
However, it is \textbf{not} a time evolution operator; it is merely used to ensure consistency, by imposing the exact chain of events that we expect to have: if the state of the external system is $\left|1\right\rangle $ at $t=1$ (time travel happens), then the corresponding state of the time machine is $\left|1\right\rangle $ at $t=0$ (the time machine is not empty), and if the state of the external system is $\left|0\right\rangle $ at $t=1$ (time travel does not happen), then the corresponding state of the time machine is $\left|0\right\rangle $ at $t=0$ (the time machine is empty).

Since the operator $T$ plays a central role in our model, let us clarify its physical meaning. A time machine is essentially an identification of two regions of spacetime, such that whatever enters it at $t=1$ is the same thing that emerges from it at $t=0$. The operator $T$ is the quantum mechanical expression of this identification. When an object passes ``through'' the time machine, there is no transit time; the two regions are identified with each other; that is, they are literally the same region, so whatever is at the entrance is also at the exit. Therefore, $T$ does not describe a process that unfolds in time, and it is not generated by a Hamiltonian. Furthermore, time travel in our model does not involve delaying, stopping, or reversing the unitary evolution of any system. All systems evolve forward in time via $U$, as usual, and $T$ merely imposes a condition relating the states at the two ends of the CTC, similar to a boundary condition.

More precisely, the action of $T$ in \eqref{eq:T-op} consists of two parts. The first part, which we described above, sets the state of the time machine at $t=0$ to the state $\left|y\right\rangle $ of the external system at $t=1$. The second part sets the state of the external system to $\left|y\mnd x\right\rangle $, which, as can be seen from \eqref{eq:unitary-U}, is simply its initial state at $t=0$. This does \textbf{not} mean that the external system travels back in time. Only the object passing through the time machine does so, while the external system continues to evolve toward the future after $t=1$. The second part of $T$ merely encodes the fact that whatever emerges from the time machine at $t=0$ always encounters the external system in its initial state, which was determined before $t=0$, and is therefore independent of anything that happens later.

The correlation operator \eqref{eq:T-op} is chosen so that the composite operator $TU$ acts as the computational basis permutation
\[
TU\left(\left|x\right\rangle \otimes\left|y\right\rangle \right)=\left|x\pld y\right\rangle \otimes\left|y\right\rangle .
\]
Since we imposed the initial condition $y=1$, the composite operator maps the time machine state as $x\mapsto x\pld1$, and maps the initial state of the external system to itself (which must be the case for consistency). For a qubit, the map $x\mapsto x\pld1$ is just a NOT gate, but in Section~\ref{subsec:Additional-timelines}, we will discuss how it generalizes to Hilbert spaces of dimension $n$.

The \textbf{consistency condition} can be written as follows:
\[
T\left|\Psi\left(1\right)\right\rangle =\left|\Psi\left(0\right)\right\rangle ,
\]
or more succinctly, since $\left|\Psi\left(1\right)\right\rangle =U\left|\Psi\left(0\right)\right\rangle $,
\[
TU\left|\Psi\left(0\right)\right\rangle =\left|\Psi\left(0\right)\right\rangle .
\]
In fact, this condition is too strong. Recall that two normalized states $\left|\Psi_{1}\right\rangle $ and $\left|\Psi_{2}\right\rangle $ are equivalent if $\left|\Psi_{1}\right\rangle =\e^{\ii\phi}\left|\Psi_{2}\right\rangle $ for some $\phi\in\left[0,2\pi\right)$. Therefore, consistency only requires the weaker condition
\[
T\left|\Psi\left(1\right)\right\rangle =\e^{\ii\phi}\left|\Psi\left(0\right)\right\rangle ,
\]
or
\begin{equation}
TU\left|\Psi\left(0\right)\right\rangle =\e^{\ii\phi}\left|\Psi\left(0\right)\right\rangle .\label{eq:consistency-condition}
\end{equation}
Let us now impose this condition and see if our system is consistent. Applying $T$ to \eqref{eq:Psi-1}, we get:
\[
T\left|\Psi\left(1\right)\right\rangle =\alpha\left|1\right\rangle \otimes\left|1\right\rangle +\beta\left|0\right\rangle \otimes\left|1\right\rangle .
\]
Comparing this to the initial state \eqref{eq:Psi-0}, we see that the consistency condition becomes
\[
\alpha\left|1\right\rangle \otimes\left|1\right\rangle +\beta\left|0\right\rangle \otimes\left|1\right\rangle =\e^{\ii\phi}\left(\alpha\left|0\right\rangle \otimes\left|1\right\rangle +\beta\left|1\right\rangle \otimes\left|1\right\rangle \right).
\]
Rearranging, we get
\[
\left(\alpha-\e^{\ii\phi}\beta\right)\left|1\right\rangle \otimes\left|1\right\rangle +\left(\beta-\e^{\ii\phi}\alpha\right)\left|0\right\rangle \otimes\left|1\right\rangle =0.
\]
This is only satisfied if $\alpha=\e^{\ii\phi}\beta$ and $\beta=\e^{\ii\phi}\alpha$. This means that $\alpha=\e^{2\ii\phi}\alpha$, so $\phi$ can be either 0 or $\pi$; taking $\phi=0$ gives the condition $\alpha=\beta$, while taking $\phi=\pi$ gives $\alpha=-\beta$. In both cases, $\left|\Psi\left(1\right)\right\rangle $ is maximally entangled. Requiring also the normalization condition $\left|\alpha\right|^{2}+\left|\beta\right|^{2}=1$, we see that both coefficients must be $1/\sqrt{2}$ in absolute value. Since the state is only defined up to an overall phase, we can, without loss of generality, take $\alpha=1/\sqrt{2}$. Therefore, the complete set of solutions is (up to an overall phase):
\[
\alpha=\frac{1}{\sqrt{2}},\qquad\beta=\pm\frac{1}{\sqrt{2}}.
\]
For simplicity, we will only consider the solution $\alpha=\beta=1/\sqrt{2}$; adding a relative minus sign will not affect the results of this paper. In conclusion, the time travel paradox is resolved by allowing the system to be in a superposition, with the states of the system at each point in time being
\[
\left|\Psi\left(0\right)\right\rangle =\frac{1}{\sqrt{2}}\left(\left|0\right\rangle +\left|1\right\rangle \right)\otimes\left|1\right\rangle ,
\]
\[
\left|\Psi\left(1\right)\right\rangle =\frac{1}{\sqrt{2}}\left(\left|0\right\rangle \otimes\left|1\right\rangle +\left|1\right\rangle \otimes\left|0\right\rangle \right).
\]

\subsection{\label{subsec:Compatibility-with-the}Compatibility with Collapse Interpretations}

We have formulated a physical system involving a time machine and shown that it is inherently inconsistent, hence paradoxical, when treated classically---but becomes fully consistent when treated quantum mechanically. However, we must now ask: is this result universal, or does it depend on interpreting quantum mechanics in a specific way?

Let us consider what happens if we allow the quantum state to collapse, whether by invoking a naive interpretation such as the Copenhagen interpretation or a more sophisticated model such as Ghirardi-Rimini-Weber (GRW) theory \cite{Ghirardi1986}. The state at $t=1$ is
\[
\left|\Psi\left(1\right)\right\rangle =\frac{1}{\sqrt{2}}\left(\left|0\right\rangle \otimes\left|1\right\rangle +\left|1\right\rangle \otimes\left|0\right\rangle \right).
\]
If we perform a measurement on $\HH_{\mathrm{ex}}$ at this time, we expect to find either $\left|0\right\rangle $ or $\left|1\right\rangle $ with probability 1/2 each. What these states actually look like depends on the specific system; in our macroscopic example, we will find Alice either dead or alive, and in our microscopic example, we will either find or not find the particle.

Since the two systems are entangled, a measurement of $\HH_{\mathrm{ex}}$ will cause the entire system $\HH_{\mathrm{CTC}}\otimes\HH_{\mathrm{ex}}$ to collapse. If the system collapsed to $\left|0\right\rangle \otimes\left|1\right\rangle $, then we have an empty time machine and a living Alice or non-annihilated particle. But if Alice puts a bomb inside the time machine, or the particle enters it, this immediately creates an inconsistency. Due to the collapse, we know for a fact that the time machine is empty with 100\% probability, so where exactly did the bomb or the particle go?

Similarly, if the system collapsed to $\left|1\right\rangle \otimes\left|0\right\rangle $, then we have a non-empty time machine and a dead Alice or annihilated particle. But if Alice is dead, then where did the bomb inside the time machine come from? And if the particle is annihilated, then where did the particle it annihilated with come from?

We conclude that our naive resolution of the paradox using quantum superposition is \textbf{not} consistent with interpretations or models involving collapse. We attempted to resolve the paradox by employing the principle of superposition, but a collapse will destroy that superposition, essentially taking us back to the classical setting and undoing the resolution we so carefully constructed.

\section{\label{sec:Many-worlds-and}Many Worlds and Entangled Timelines}

In the previous section, we showed how our time travel paradox is resolved by classical parallel timelines. However, we did not provide any mechanism for generating these timelines. We also showed that quantum superposition can resolve the paradox, but this is not compatible with collapse interpretations. Fortunately, we can solve both issues by adopting the \textbf{Everett or ``many-worlds'' interpretation (MWI)} of quantum mechanics \cite{Everett1957,DeWitt1970,DeWitt1973,Deutsch1985,Wallace2003,Saunders2010,Everett2012,Wallace2012,Vaidman2024}. Let us first review how this interpretation works and clarify some subtleties and misconceptions.

\subsection{\label{subsec:Quantum-mechanics-without}Quantum Mechanics Without Collapse}

Imagine that Alice has a qubit in some superposition of $\left|0\right\rangle $ and $\left|1\right\rangle $:
\[
\left|\textrm{qubit}\right\rangle =a\left|0\right\rangle +b\left|1\right\rangle ,\qquad\left|a\right|^{2}+\left|b\right|^{2}=1.
\]
Upon measuring the qubit in the computational basis $\left\{ \left|0\right\rangle ,\left|1\right\rangle \right\} $, Alice will measure 0 with probability $\left|a\right|^{2}$ or 1 with probability $\left|b\right|^{2}$. Collapse interpretations of quantum mechanics then assume there is some \textbf{non-unitary} evolution whereby the state ``collapses'' into either $\left|0\right\rangle $ or $\left|1\right\rangle $ based on the value that was measured. This contradicts the requirement that the evolution of a closed quantum system be described by a \textbf{unitary} operator \cite{Shoshany2018}.

To resolve this issue, the MWI suggests that we should take into account not just the state of the qubit but also Alice's state. Before the measurement, Alice is in the initial state $\left|\textrm{Alice}\right\rangle $, which can be roughly defined as the composite state, at that point in time, of the various quantum systems Alice is composed of. The composite state $\left|\Psi\left(0\right)\right\rangle $ of the joint qubit + Alice system before the measurement, at time $t=0$, is
\[
\left|\Psi\left(0\right)\right\rangle =\left(a\left|0\right\rangle +b\left|1\right\rangle \right)\otimes\left|\textrm{Alice}\right\rangle .
\]
Notice that this is a separable state. After the measurement, at time $t=1$, the system undergoes unitary evolution as follows:
\begin{equation}
\left|\Psi\left(1\right)\right\rangle =a\left|0\right\rangle \otimes\left|\textrm{Alice measured 0}\right\rangle +b\left|1\right\rangle \otimes\left|\textrm{Alice measured 1}\right\rangle .\label{eq:MWI-branching}
\end{equation}
The operator performing this evolution is similar to a CNOT gate \cite{Nielsen2010}: it essentially checks the state of the qubit and changes Alice's state accordingly. It is unitary, and thus invertible and preserves probabilities. This is in contrast with the collapse operation, which is not unitary. Thus, the MWI is sometimes referred to as \textbf{purely unitary quantum mechanics}; it is simply quantum mechanics without the additional assumption of non-unitary collapse.

Importantly, if $a$ and $b$ are both nonzero, the states of the qubit and Alice are now entangled. As the measurement takes place, photons, air molecules, and other environmental degrees of freedom also become correlated with the measurement outcome. Let $\left|E_{0}\right\rangle $ and $\left|E_{1}\right\rangle $ denote the corresponding states of the environment, and define $\left|S_{0}\right\rangle \equiv\left|0\right\rangle \otimes\left|\textrm{Alice measured 0}\right\rangle $ and $\left|S_{1}\right\rangle \equiv\left|1\right\rangle \otimes\left|\textrm{Alice measured 1}\right\rangle $ for brevity. Then the joint state of the qubit, Alice, and the environment is
\[
\left|\Psi_{SE}\right\rangle =a\left|S_{0}\right\rangle \otimes\left|E_{0}\right\rangle +b\left|S_{1}\right\rangle \otimes\left|E_{1}\right\rangle .
\]
The reduced state of the qubit + Alice system is obtained by tracing out the environment:
\[
\begin{aligned}\rho_{S} & \equiv\tr_{E}\left(\left|\Psi_{SE}\right\rangle \left\langle \Psi_{SE}\right|\right)\\
 & =\left|a\right|^{2}\left|S_{0}\right\rangle \left\langle S_{0}\right|+\left|b\right|^{2}\left|S_{1}\right\rangle \left\langle S_{1}\right|+ab^{*}\left\langle E_{1}|E_{0}\right\rangle \left|S_{0}\right\rangle \left\langle S_{1}\right|+a^{*}b\left\langle E_{0}|E_{1}\right\rangle \left|S_{1}\right\rangle \left\langle S_{0}\right|.
\end{aligned}
\]
Interactions with numerous environmental degrees of freedom make $\left|E_{0}\right\rangle $ and $\left|E_{1}\right\rangle $ effectively orthogonal, $\left\langle E_{0}|E_{1}\right\rangle \approx0$, and thus the off-diagonal terms in $\rho_{S}$ are suppressed, even though $\left|\Psi_{SE}\right\rangle $ remains pure. This is the process of \textbf{decoherence} \cite{Zurek2003b,Schlosshauer2007}, which was not part of Everett's original formulation but plays an essential role in the modern understanding of the MWI. Note that decoherence does not eliminate the superposition. Both terms are still present in $\left|\Psi_{SE}\right\rangle $, and the environment is itself part of the entangled state, with a different state in each term. What decoherence suppresses is the interference between the terms, so that each of them evolves, for all practical purposes, independently of the other.

We interpret this entanglement and the resulting decoherence as a \textbf{branching}: the overall quantum state of the qubit and Alice has transitioned from a separable state describing each system separately to an entangled state describing both systems as being inseparably correlated.

The states $\left|\textrm{Alice measured 0}\right\rangle $ and $\left|\textrm{Alice measured 1}\right\rangle $ are two physically distinct states, as Alice's brain (which is a subset of the quantum systems Alice is composed of) registered a different measurement result in each. We interpret this as Alice herself branching into two versions, one who measured 0 and another who measured 1. These two versions then continue living their lives separately, as if in two different ``worlds''.

However, these ``worlds'' are not physically distinct universes but rather just two different terms in the superposition of the overall system. In the MWI, a superposition state like \eqref{eq:MWI-branching} does not represent possible measurement outcomes or an observer's lack of knowledge about the system; it is interpreted as representing a multiplicity of ``worlds'' which are emergent structures inferred from the overall quantum state \cite{Wallace2003,Wallace2012}.

How is this approach compatible with the experimental fact that systems do \textbf{seem} to collapse into a single state upon measurement? The answer is that decoherence makes the other branches invisible to observers within each branch. The version of Alice in the branch corresponding to the term $\left|S_{0}\right\rangle \otimes\left|E_{0}\right\rangle $ will conclude that the qubit collapsed to the state $\left|0\right\rangle $ \textbf{from her perspective}, because she has no access to the branch where the qubit is in the state $\left|1\right\rangle $, and vice versa.

\subsection{\label{subsec:Some-subtleties}Some Subtleties}

In this example, we treated Alice as if she were a simple quantum system with 3 states: $\left|\textrm{Alice}\right\rangle $, $\left|\textrm{Alice measured 0}\right\rangle $, and $\left|\textrm{Alice measured 1}\right\rangle $, which are eigenstates of some Hermitian ``Alice measurement'' operator $A$. However, it should be stressed that this is obviously not very precise. It is slightly more precise to say that each of these eigenstates labels an \textbf{equivalence class} of states with a similar macroscopic interpretation.

For example, the state $\left|\textrm{Alice measured 0}\right\rangle $ tells us that Alice measured 0 for the qubit, but there is actually an enormous---possibly infinite---number of states that can be interpreted in this way. Is Alice sitting or standing? Is she breathing in or out? What is the quantum state of a certain carbon atom in her nose?

The possible answers to these questions correspond to different states in Alice's Hilbert space, but the distinction between these states is not important for our purposes, so we collect all of them into one equivalence class labeled by $\left|\textrm{Alice measured 0}\right\rangle $. As long as the neurons in Alice's brain registered that the qubit's value is 0, Alice's state will belong to this equivalence class.

To be even more precise, we must assume that any ``in-between'' states have sufficiently low probability that they can be ignored. For example, one can also imagine a situation where the measuring apparatus malfunctioned and displayed the value 2. This does not count as a state in the equivalence classes labeled by $\left|\textrm{Alice measured 0}\right\rangle $ or $\left|\textrm{Alice measured 1}\right\rangle $, but we ignore it in our discussion just as we ignore the possibility that a coin lands on its side when we consider a coin toss.

The state \eqref{eq:MWI-branching} should therefore be interpreted more precisely not as a state with exactly two distinct branches but rather as a coarse-grained description of numerous (probably infinitely many) branches, which belong to the two main equivalence classes labeled by $\left|\textrm{Alice measured 0}\right\rangle $ and $\left|\textrm{Alice measured 1}\right\rangle $, plus additional ``in-between'' states with very low probability. For our purposes, it does not matter which specific state from the equivalence class the system is actually in, since we only care about one specific detail: whether Alice measured 0 or 1.

\subsection{\label{subsec:Spreading-of-branches}Spreading of Branches}

How does a term in a superposition of only two entangled systems become its own ``world''? Let us consider Alice's friend, Bob, who asks Alice what the result of her measurement was. Before asking for that information, Bob is in the initial state $\left|\textrm{Bob}\right\rangle $. The composite state $\left|\Phi\left(0\right)\right\rangle $ of the joint qubit + Alice + Bob system before the measurement, at time $t=0$, is
\[
\left|\Phi\left(0\right)\right\rangle =\left(a\left|0\right\rangle +b\left|1\right\rangle \right)\otimes\left|\textrm{Alice}\right\rangle \otimes\left|\textrm{Bob}\right\rangle .
\]
After the measurement, at time $t=1$, it becomes
\[
\left|\Phi\left(1\right)\right\rangle =\left(a\left|0\right\rangle \otimes\left|\textrm{Alice measured 0}\right\rangle +b\left|1\right\rangle \otimes\left|\textrm{Alice measured 1}\right\rangle \right)\otimes\left|\textrm{Bob}\right\rangle .
\]
Note that if $a$ and $b$ are both nonzero, the qubit and Alice are now entangled, but Bob is still an independent system. However, after Bob receives the result of the measurement from Alice, at time $t=2$, the system evolves to
\begin{align*}
\left|\Phi\left(2\right)\right\rangle = & a\left|0\right\rangle \otimes\left|\textrm{Alice measured 0}\right\rangle \otimes\left|\textrm{Bob received 0}\right\rangle \\
 & +b\left|1\right\rangle \otimes\left|\textrm{Alice measured 1}\right\rangle \otimes\left|\textrm{Bob received 1}\right\rangle .
\end{align*}
Now \textbf{all three systems} are entangled. This means that the branching has \textbf{spread} from the qubit + Alice system to the larger qubit + Alice + Bob system. In the same manner, the entanglement will continue to spread, via purely unitary evolution, throughout the causal future of the interaction. In the limit $t\to\infty$, the joint state of all systems in this causal future will be a superposition of two branches: one where the qubit had the value 0, and another where it had the value 1.

Thus the branching of worlds is not a global event but rather a \textbf{local branching} which spreads causally outwards from the spacetime location of the initial measurement or interaction. This fact is worth emphasizing, as it is often misunderstood in casual discussions of the MWI. The two most common misconceptions resulting from this misunderstanding are as follows.

First, branching does not only happen when something is measured; it happens whenever interacting systems become entangled with each other and decohere through entanglement with their environment. In our example, Alice performs a ``measurement'' of the state of the qubit, but in reality, Alice simply interacts with the qubit, or perhaps with some apparatus which in turn interacts with the qubit (as in a Stern-Gerlach experiment to measure the spin of an electron); a ``measurement'' does not have any special properties that other types of interaction do not have.

Indeed, in the second step of our scenario, when Bob asks Alice about the result of her measurement, there is no conventional ``measurement'' happening but rather a simple exchange of information. However, this step is exactly the same as the first step, if we treat the qubit + Alice system as a single system in a superposition.

Second, and more importantly, the measurement or interaction itself does not ``create'' physically new universes. It merely causes two systems to become entangled, and this entanglement spreads out as more systems interact with the initial system and become entangled with it in turn.

In this process there is always only \textbf{one} universal quantum state, not multiple physically distinct universes, although the term ``multiverse'' is often used in the literature to describe its emergent branch structure. We are merely considering the entanglement between subsystems in a very large (possibly infinite) composite Hilbert space. The decohered quasi-classical components of the universal quantum state are interpreted as effectively independent ``worlds'', each corresponding to a particular outcome of the initial measurement.

Rather than ``many worlds'' or ``multiverse'', a more precise term could perhaps be ``entangled worlds'' or ``entangled histories''. As we will show, in our model, the ``worlds'' become different timelines and can be used to resolve time travel paradoxes. Therefore, we decided to name our model \textbf{entangled timelines}, to stress that the timelines are not entirely new universes that are somehow created out of nothing (as in the naive ``branching timelines'' picture) but rather the result of entanglement with the time machine followed by decoherence. The meaning of this will become clearer in the next section.

\subsection{\label{subsec:Resolving-the-paradox}Resolving the Paradox with Entangled Timelines}

We now apply the MWI to our time travel paradox model. Recall the states of the system at $t=0$ and $t=1$ from Section~\ref{subsec:Resolution-using-quantum}:
\begin{equation}
\left|\Psi\left(0\right)\right\rangle =\frac{1}{\sqrt{2}}\left(\left|0\right\rangle +\left|1\right\rangle \right)\otimes\left|1\right\rangle ,\label{eq:Psi-0-2}
\end{equation}
\begin{equation}
\left|\Psi\left(1\right)\right\rangle =\frac{1}{\sqrt{2}}\left(\left|0\right\rangle \otimes\left|1\right\rangle +\left|1\right\rangle \otimes\left|0\right\rangle \right).\label{eq:Psi-1-2}
\end{equation}
Recall also that in each tensor product, the state on the left represents the time machine $\HH_{\mathrm{CTC}}$ and the state on the right represents the external system $\HH_{\mathrm{ex}}$.

The state at $t=0$ is separable. At this point, time travel has not yet happened and the external system has not yet interacted with the time machine. Alice has not yet opened the door to see if there is a bomb inside, or the particle has not yet reached the point where it might annihilate with its future self. This joint state is thus similar to that of Alice and the qubit in our example from Section~\ref{subsec:Quantum-mechanics-without}, before they interacted. We assume, for simplicity, that the time machine remains isolated from its environment until it interacts with the external system.

At $t=0$, no branching has yet occurred. The time machine is in a coherent superposition of being empty ($\left|0\right\rangle $) and not empty ($\left|1\right\rangle $), while the external system remains in its initial state $\left|1\right\rangle $.

At $t=1$, after the interaction, the states of the time machine and the external system become entangled. Again, this is similar to Alice becoming entangled with the state of the qubit in Section~\ref{subsec:Quantum-mechanics-without}. Once these branches have decohered, an observer within the external system will thus be split into two versions of themselves. One version sees that the time machine is empty and no interaction happened, while the other version sees that the time machine is not empty and an interaction has occurred which prevented time travel from happening.

In the context of the MWI, where the branches correspond to decohered records of possible outcomes in a measurement, we interpret each branch as an apparent ``collapse'' for the observer within that branch. However, in the context of time travel, we interpret each branch as a separate timeline, where in the first branch no time travel has occurred, while in the second branch time travel did occur and the local history is different.

Each timeline evolves from $t=0$ to $t=1$ using the unitary evolution operator $U$ given by \eqref{eq:unitary-U}. Note that this operator, which has the same form as a CNOT gate, entangles the state; this encodes the fact that the interaction between the time machine $\HH_{\mathrm{CTC}}$ and the external system $\HH_{\mathrm{ex}}$ entangles the two systems and, after decoherence, ``spreads'' the timeline to the external system.

The timeline correlation operator $T$ given by \eqref{eq:T-op}, when applied at $t=1$, tells us how to go from one timeline to the next: it will take us from the first term in the superposition at $t=1$ to the second term in the superposition at $t=0$, and then from the second term in the superposition at $t=1$ back to the first term in the superposition at $t=0$. Thus, the two operators $U$ and $T$ allow us to trace the evolution within each individual timeline (using $U$) as well as the relationships between the timelines themselves (using $T$).

Since this is the central part of our model, let us describe in more detail what happens physically, using the macroscopic example of Section~\ref{subsec:A-macroscopic-example} for concreteness. At $t=0$, the state of the time machine and Alice is
\[
\frac{1}{\sqrt{2}}\left(\left|\textrm{empty}\right\rangle +\left|\textrm{bomb}\right\rangle \right)\otimes\left|\textrm{alive}\right\rangle .
\]
 At this time, there is only one version of Alice; only the time machine is in a superposition. Alice then opens the door of the time machine. As she interacts with its contents, the state evolves into
\[
\frac{1}{\sqrt{2}}\left(\left|\textrm{empty}\right\rangle \otimes\left|\textrm{alive}\right\rangle +\left|\textrm{bomb}\right\rangle \otimes\left|\textrm{dead}\right\rangle \right).
\]
 The two terms are the two timelines, $h=0$ and $h=1$, each containing its own version of Alice.

At $t=1$, Alice of timeline $h=0$ is alive and places a bomb inside the time machine, while Alice of timeline $h=1$ is dead, so nothing is placed inside it. Since whatever enters the time machine at $t=1$ is the same thing that emerges from it at $t=0$, the term $\left|\textrm{bomb}\right\rangle $ in the superposition at $t=0$ is the bomb sent by Alice of timeline $h=0$, while the term $\left|\textrm{empty}\right\rangle $ corresponds to nothing being sent in timeline $h=1$. This is precisely the content of the operator $T$:
\[
T\left(\left|\textrm{empty}\right\rangle \otimes\left|\textrm{alive}\right\rangle \right)=\left|\textrm{bomb}\right\rangle \otimes\left|\textrm{alive}\right\rangle ,\qquad T\left(\left|\textrm{bomb}\right\rangle \otimes\left|\textrm{dead}\right\rangle \right)=\left|\textrm{empty}\right\rangle \otimes\left|\textrm{alive}\right\rangle .
\]
 Note that the timelines are parallel, not consecutive. Both of them span the same time interval, and the cycle formed by $T$ is a cycle of correlations between the timelines, not a repetition of events in time.

It is important to stress that this does not require the two timelines to merge or interfere with each other, and in particular, it does not require any process that reverses decoherence. The operator $T$ is a permutation of the basis states, so it acts on each term in the superposition separately, relating it to a term in the superposition at $t=0$, when the time machine has not yet interacted with Alice or with the environment. In other words, $t=0$ is the point from which the timelines begin to diverge. The bomb sent from timeline $h=0$ does not merge with the earlier contents of the time machine; rather, it \textbf{is} the contents of the time machine at $t=0$ in the term $\left|\textrm{bomb}\right\rangle $ of the superposition, and timeline $h=1$ is what develops from this term once Alice opens the door.

Note also that Alice is not confined to a closed loop in time. Only the object passing through the time machine travels along the CTC. Alice arrives from the past, interacts with the time machine between $t=0$ and $t=1$, and then continues toward the future, in both timelines (although in timeline $h=1$, she does so in the state $\left|\textrm{dead}\right\rangle $). Hence, the region around the time machine is not causally disconnected from the rest of the universe. Indeed, as we will see in Section~\ref{subsec:Spreading-of-the}, the timelines subsequently spread to any other system which interacts with it.

Finally, let us comment on the role of the environment. In our model, the time machine and the external system, including all of their microscopic degrees of freedom, are assumed to be isolated from their environment until $t=1$, so that $T$ acts on a pure state. Decoherence by the environment takes place afterwards, as they interact with it, and each timeline then contains its own version of the environment as well, as we will see in Section~\ref{subsec:Spreading-of-the}. The consistency condition \eqref{eq:consistency-condition} was solved in Section~\ref{subsec:Correlation-and-consistency} for the time machine and the external system alone, and its solution does not depend on the environment in any way. Therefore, the paradox is resolved even if the environment is finite, such that decoherence is incomplete, or absent altogether, as may be the case in the microscopic example of Section~\ref{subsec:A-microscopic-example}. What does depend on the environment is the interpretation of the terms in the superposition as timelines. Without decoherence, the two terms can still interfere with each other in principle, and the decomposition of the state into these two particular terms is not preferred over any other. The system is then simply in an entangled superposition, like any other isolated quantum system, and the notion of ``timelines'' is no more meaningful than the notion of ``worlds'' in the MWI before decoherence has occurred. For a macroscopic system such as Alice, on the other hand, decoherence is extremely fast and practically irreversible. Even if Alice were completely isolated, her own microscopic degrees of freedom would serve as an environment. Moreover, in general the environment is not limited to a finite region, as it includes any system in the causal future of the interaction.

The same caveats we discussed in Section~\ref{subsec:Some-subtleties} still apply here: to be more precise, we should interpret the state \eqref{eq:Psi-1-2} as a coarse-grained description of numerous (probably infinitely many) timelines belonging to the two main equivalence classes labeled by $\left|0\right\rangle \otimes\left|1\right\rangle $ and $\left|1\right\rangle \otimes\left|0\right\rangle $, plus additional ``in-between'' states (e.g., where the time machine malfunctioned) with very low probability. For our purposes, it does not matter which specific state from the equivalence class the system is actually in, since we only care whether the time machine is empty or not, and whether time travel will happen or not.

These caveats certainly apply to the macroscopic paradox of Section~\ref{subsec:A-macroscopic-example}, where Alice's states $\left|\textrm{alive}\right\rangle $ and $\left|\textrm{dead}\right\rangle $, for example, are really labels for equivalence classes of a myriad of different states corresponding to Alice being alive or dead, respectively. However, they also apply to the microscopic paradox of Section~\ref{subsec:A-microscopic-example} since, for example, the states $\left|\textrm{annihilated}\right\rangle $ and $\left|\textrm{not annihilated}\right\rangle $ of the particle are likewise labels for equivalence classes of a myriad of states of the quantum field describing said particle, corresponding to the particle being either annihilated or not annihilated, respectively.

\subsection{The Role of Probabilities}

Let us now consider the interpretation of the coefficients in the superpositions \eqref{eq:Psi-0-2} and \eqref{eq:Psi-1-2} as probability amplitudes. In Section~\ref{subsec:Correlation-and-consistency} we started with two arbitrary coefficients $\alpha$ and $\beta$ and proved that the evolution is only consistent (and the paradox resolved) if $\alpha=\pm\beta$. This means that the probability associated with each timeline is 1/2. But why would that be the case?

Logically, it is easy to see why this is the only consistent option. Indeed, considering for example the microscopic paradox of Section~\ref{subsec:A-microscopic-example}, imagine that the particle annihilates with probability 1/3 and does not annihilate with probability 2/3. But no annihilation leads to a particle coming out of the time machine and causing an annihilation, so annihilation should \textbf{also} happen with probability 2/3, creating a contradiction.

However, let us also present another argument. One of the most common objections to the MWI is the \textbf{probability problem}. If every outcome simply corresponds to a different branch of the universal quantum state, why would an observer measuring the state $\sqrt{1/3}\left|0\right\rangle +$$\sqrt{2/3}\left|1\right\rangle $ measure 0 a third of the time and 1 two thirds of the time, if each outcome has exactly one branch associated with it?

One proposed solution to the problem is that the probabilities are associated with the number of identical branches \cite{Aharonov2005} (Section 9.3). For example, in the state $\sqrt{1/3}\left|0\right\rangle +$$\sqrt{2/3}\left|1\right\rangle $, there will be not two but \textbf{three} branches, where one branch corresponds to 0 and the other two correspond to 1. Unfortunately, this solution is flawed, as it merely ``moves the goalposts'', and does not explain how the coefficients of the superposition can be related to the number of branches, or why two identical branches should count as two and not just one.

Nevertheless, this solution appears to be perfectly tailored to the case of entangled timelines. Clearly, we have exactly two distinct timelines in our model, one where the time machine is empty and one where it is not empty, and therefore, two branches. If we take the suggestion that the probabilities come from the number of branches seriously, it is immediately clear that the corresponding probabilities \textbf{must} be 1/2. More generally, in a model with $n$ timelines, the probability for each timeline must be $1/n$, as we will show explicitly in Section~\ref{subsec:Additional-timelines}.

The role of probabilities and the Born rule in the MWI is an area of active research, and further discussion is beyond the scope of our paper, which merely makes use of the well-established MWI formalism to resolve time travel paradoxes. We refer the reader to Refs.~\cite{Deutsch1999,Wallace2012,Sebens2018,McQueen2019,Vaidman2021} for discussions of several different approaches.

\subsection{\label{subsec:Dependence-on-the}Dependence on the Interpretation}

Since the MWI is far from universally accepted, it is important to clarify the extent to which our results depend on it. The formalism of the E-CTC model does not depend on any particular interpretation. The states of the system, the operators $U$ and $T$, the consistency condition \eqref{eq:consistency-condition}, and its solutions were all obtained in Sections \ref{subsec:Resolution-using-quantum} and \ref{subsec:Correlation-and-consistency} using nothing but the standard formalism of quantum mechanics, before the MWI was even introduced. Furthermore, the magnitudes of the coefficients in the superposition were completely determined by the consistency condition alone, without making use of the Born rule. Therefore, the consistency of the model does not require first resolving the probability problem discussed in the previous section.

What does depend on the interpretation is whether these solutions can be understood as resolving the paradox using parallel timelines. For this, two requirements must be satisfied. First, the superposition must not be destroyed by a physical collapse process, as we showed in Section~\ref{subsec:Compatibility-with-the}. Second, all the terms in the superposition must be equally real, since each timeline is the origin of whatever emerges from the time machine in the next timeline. If only one of the terms describes what actually happens, then we are back to a single timeline, and the paradox returns.

The MWI is the simplest framework satisfying both requirements, which is why we adopted it in this paper. However, the model does not rely on any feature of the MWI beyond these two requirements. Other approaches in which the quantum state always evolves unitarily, such as relational \cite{Rovelli1996} or perspectival \cite{Dieks2019} interpretations, where the state of a system is only defined relative to another system, may be able to accommodate our model as well. The timelines would then be understood not as ``worlds'' but as different perspectives, each defined relative to a different state of the external system. We leave this question to future work.

\subsection{\label{subsec:Spreading-of-the}Spreading of the Timelines}

In our model, two timelines are created when the external system $\HH_{\mathrm{ex}}$ interacts with the time machine and decoherence renders the resulting components effectively independent. As with the branches of the MWI, the timelines do not instantaneously spring into existence everywhere. Other systems interacting with the $\HH_{\mathrm{CTC}}\otimes\HH_{\mathrm{ex}}$ system then get ``infected'', so to speak, with the split into timelines. The timelines will thus spread gradually throughout the causal future of the interaction, as described in Section~\ref{subsec:Spreading-of-branches} for the MWI.

To illustrate this, consider the microscopic example of Section~\ref{subsec:A-microscopic-example}. We place the entire $\HH_{\mathrm{CTC}}\otimes\HH_{\mathrm{ex}}$ system inside a box, and place a particle detector outside the box. In timeline $h=0$, where no annihilation occurred and the particle went into the time machine, the detector will not detect any particles outside the box. However, in timeline $h=1$, where annihilation did occur, the detector will register the final products of the annihilation, which for concreteness we will assume here to be one or more photons, outside the box.

We now consider the composite system of the box and the detector, similar to the composite system qubit + Alice + Bob we considered in Section~\ref{subsec:Spreading-of-branches}. The detector can be in one of two states: $\left|0\right\rangle $ if no photon was detected or $\left|1\right\rangle $ if a photon was detected. At time $t=0$, this enlarged system will be in the state
\[
\left|\Psi\left(0\right)\right\rangle =\frac{1}{\sqrt{2}}\left(\left|\textrm{empty}\right\rangle +\left|\textrm{particle}\right\rangle \right)\otimes\left|\textrm{not annihilated}\right\rangle \otimes\left|0\right\rangle ,
\]
which is completely separable; no two systems are entangled. At time $t=1$, after the time machine and the external system interacted (possibly leading to particle annihilation), the composite system will be in the state
\[
\left|\Psi\left(1\right)\right\rangle =\frac{1}{\sqrt{2}}\left(\left|\textrm{empty}\right\rangle \otimes\left|\textrm{not annihilated}\right\rangle +\left|\textrm{particle}\right\rangle \otimes\left|\textrm{annihilated}\right\rangle \right)\otimes\left|0\right\rangle .
\]
The time machine and external system are now entangled, and after decoherence we have two timelines \textbf{locally inside the box}. However, the detector is not yet entangled with the other two systems, so it has not yet been split into two timelines.

At $t=2$, after a photon either passed or did not pass through the detector, the state will be
\[
\left|\Psi\left(2\right)\right\rangle =\frac{1}{\sqrt{2}}\left(\left|\textrm{empty}\right\rangle \otimes\left|\textrm{not annihilated}\right\rangle \otimes\left|0\right\rangle +\left|\textrm{particle}\right\rangle \otimes\left|\textrm{annihilated}\right\rangle \otimes\left|1\right\rangle \right).
\]
Now the detector is entangled with the time machine and the external system, so we conclude that the detector has now \textbf{also} split into two timelines---one timeline where no photon was detected, and another where a photon was detected. Instead of being localized to the box, the timelines have spread out to include the detector. It is now a simple matter to consider the timelines spreading further to the person reading the result on the detector's screen, and so on.

Similarly, in the macroscopic example of Section~\ref{subsec:A-macroscopic-example}, we now consider Alice's friend, Bob, approaching the lab (containing Alice and the time machine) in the initial state $\left|\textrm{happy}\right\rangle $. At $t=0$, before Alice opens the time machine, the state of the system (including Bob) will be
\[
\left|\Psi\left(0\right)\right\rangle =\frac{1}{\sqrt{2}}\left(\left|\textrm{empty}\right\rangle +\left|\textrm{bomb}\right\rangle \right)\otimes\left|\textrm{alive}\right\rangle \otimes\left|\textrm{happy}\right\rangle .
\]
At time $t=1$, after Alice opens the time machine but before Bob opens the door to the lab, the state will be
\[
\left|\Psi\left(1\right)\right\rangle =\frac{1}{\sqrt{2}}\left(\left|\textrm{empty}\right\rangle \otimes\left|\textrm{alive}\right\rangle +\left|\textrm{bomb}\right\rangle \otimes\left|\textrm{dead}\right\rangle \right)\otimes\left|\textrm{happy}\right\rangle .
\]
After decoherence, Alice has split into two timelines, but Bob is not yet entangled with the lab, so he has not yet split. Finally, at $t=2$, Bob opens the door to the lab. In one timeline, Bob finds Alice alive and remains happy, but in the other timeline, he finds her dead and is shocked:
\[
\left|\Psi\left(2\right)\right\rangle =\frac{1}{\sqrt{2}}\left(\left|\textrm{empty}\right\rangle \otimes\left|\textrm{alive}\right\rangle \otimes\left|\textrm{happy}\right\rangle +\left|\textrm{bomb}\right\rangle \otimes\left|\textrm{dead}\right\rangle \otimes\left|\textrm{shocked}\right\rangle \right).
\]
Bob has now become entangled with the lab, and therefore the timelines have spread to him as well. Next, Bob can call Alice's parents to let them know what happened, in which case the timelines spread farther out to the parents, and so on. We see that the time machine has caused a local split into two timelines, which then spreads out in space as the information is shared with more and more observers.

\subsection{Bootstrap Paradoxes}

Let us now show that our model can also resolve bootstrap paradoxes. Imagine, for example, the following scenario, first described in the introduction:
\begin{itemize}
\item At $t=0$, Bob, struggling with writer's block, opens the door to the time machine and finds a finished book inside. He publishes the book and becomes a best-selling author.
\item At $t=1$, Bob puts the book inside the time machine in order to avoid a consistency paradox.
\end{itemize}

While this scenario is perfectly consistent, one might worry that the book was seemingly created from nothing. However, this worry can be easily resolved using parallel timelines.

In timeline $h=0$, Bob opens the door to the time machine at $t=0$ and finds it empty. Disappointed, he then continues the hard work of writing his book, and eventually finishes it. At $t=1$, he puts the book inside his time machine and sends it back to $t=0$. The book arrives at $t=0$ in timeline $h=1$, and is received by a happy Bob, who then profits from the work of his clone from $h=0$ without having to do any of the work himself.

Let us describe this scenario more precisely using the entangled timelines model. The initial state at $t=0$, right before Bob opens the door to the time machine, is
\[
\left|\Psi\left(0\right)\right\rangle =\left(\alpha\left|\textrm{empty}\right\rangle +\beta\left|\textrm{book}\right\rangle \right)\otimes\left|\textrm{writing}\right\rangle ,
\]
where the state $\left|\textrm{writing}\right\rangle $ means Bob is currently working hard on writing his book. At this point, Bob has not yet interacted with the time machine. He then opens the door and becomes entangled with the contents of the time machine via the action of a unitary evolution operator $U$ analogous to \eqref{eq:unitary-U}:
\[
\left|\Psi\left(1\right)\right\rangle =\alpha\left|\textrm{empty}\right\rangle \otimes\left|\textrm{writing}\right\rangle +\beta\left|\textrm{book}\right\rangle \otimes\left|\textrm{not writing}\right\rangle .
\]
After decoherence, the two terms represent separate timelines. Here, $\left|\textrm{writing}\right\rangle $ means Bob did not find the book and is therefore continuing to write (just as Alice continues to be alive if she did not find the bomb, or the particle continues to be non-annihilated if another particle does not emerge from the time machine). The state $\left|\textrm{not writing}\right\rangle $, on the other hand, means Bob found the book and can now enjoy its profits without working hard on writing it.

We still need to show consistency. In the first timeline (with coefficient $\alpha$), Bob places the book inside the time machine. Therefore, this timeline leads, via the action of a timeline correlation operator $T$ analogous to \eqref{eq:T-op}, to the state $\left|\textrm{book}\right\rangle $ at $t=0$. However, in the other timeline (with coefficient $\beta$), Bob does not place the book inside the time machine. Therefore, this timeline leads, again via $T$, to the state $\left|\textrm{empty}\right\rangle $ at $t=0$. Bob's state at $t=0$, meanwhile, must always be the initial state $\left|\textrm{writing}\right\rangle $. We therefore get:
\[
\left|\Psi\left(0\right)\right\rangle =\alpha\left|\textrm{book}\right\rangle \otimes\left|\textrm{writing}\right\rangle +\beta\left|\textrm{empty}\right\rangle \otimes\left|\textrm{writing}\right\rangle .
\]
As usual, the only way to make this consistent is to take $\alpha=\pm\beta$ with $\left|\alpha\right|=\left|\beta\right|=1/\sqrt{2}$.

\subsection{\label{subsec:Additional-timelines}Additional Timelines}

So far, for simplicity, we have only considered scenarios with two timelines. However, it is a straightforward matter to add more timelines. Let us consider again our generic paradox, which has the states
\begin{equation}
\left|\Psi\left(0\right)\right\rangle =\left(\alpha\left|0\right\rangle +\beta\left|1\right\rangle \right)\otimes\left|1\right\rangle ,\label{eq:n-timelines-2-t0}
\end{equation}
\begin{equation}
\left|\Psi\left(1\right)\right\rangle =\alpha\left|0\right\rangle \otimes\left|1\right\rangle +\beta\left|1\right\rangle \otimes\left|0\right\rangle .\label{eq:n-timelines-2-t1}
\end{equation}
Remember that each timeline corresponds to a different time machine state. Therefore, for $n$ different timelines, let us define $\HH_{\mathrm{CTC}}$ as an $n$-dimensional Hilbert space with an orthonormal basis $\left|i\right\rangle $ where $i\in\left\{ 0,\ldots,n-1\right\} $, and choose a generic superposition with amplitudes $\alpha_{i}$. Then the initial state of the time machine is
\[
\left|\Psi_{\mathrm{CTC}}\left(0\right)\right\rangle =\sum^{n-1}_{i=0}\alpha_{i}\left|i\right\rangle ,
\]
a superposition of $n$ states, one for each possible timeline. The coefficients must satisfy the normalization condition
\begin{equation}
\sum^{n-1}_{i=0}\left|\alpha_{i}\right|^{2}=1.\label{eq:n-normalization}
\end{equation}
Recall also that the timelines emerge from the interaction between the external system and the time machine through entanglement and decoherence. Therefore, we also take the external Hilbert space $\HH_{\mathrm{ex}}$ to be $n$-dimensional, with orthonormal basis states $\left|i\right\rangle $ where $i\in\left\{ 0,\ldots,n-1\right\} $. This allows the construction below to produce a state with Schmidt rank $n$.

As usual, we take the initial state of $\HH_{\mathrm{ex}}$ to be $\left|1\right\rangle $, and this state is the same in all timelines, as it is an initial condition determined before time travel takes place. So at $t=0$ we have
\begin{equation}
\left|\Psi\left(0\right)\right\rangle =\left(\sum^{n-1}_{i=0}\alpha_{i}\left|i\right\rangle \right)\otimes\left|1\right\rangle .\label{eq:n-timelines}
\end{equation}
Note that this reduces to \eqref{eq:n-timelines-2-t0} for $n=2$ with $\alpha_{0}=\alpha$ and $\alpha_{1}=\beta$.

The unitary evolution operator $U$ defined in \eqref{eq:unitary-U} readily generalizes to the case of $n$ timelines:
\[
U\left(\left|x\right\rangle \otimes\left|y\right\rangle \right)=\left|x\right\rangle \otimes\left|x\pld y\right\rangle ,
\]
where now $\pld$ represents addition modulo $n$. After we apply this operator, we find that the state at $t=1$ is
\[
\left|\Psi\left(1\right)\right\rangle =U\left|\Psi\left(0\right)\right\rangle =\sum^{n-1}_{i=0}\alpha_{i}\left|i\right\rangle \otimes\left|i\pld1\right\rangle .
\]
As expected, the state is entangled (whenever more than one coefficient is nonzero). After decoherence, we interpret each term as corresponding to a different timeline. Again, this reduces to \eqref{eq:n-timelines-2-t1} for $n=2$ with $\alpha_{0}=\alpha$ and $\alpha_{1}=\beta$.

In the first timeline ($i=0$) the external system remains in its initial state $\left|1\right\rangle $, as in that timeline no interaction occurred. However, in subsequent timelines ($0<i<n-1$) it evolves to the state $\left|i\pld1\right\rangle $ due to interaction with the time machine. Finally, in the last timeline ($i=n-1$) the external system evolves to the state $\left|0\right\rangle $, which prevents any further time travel from occurring.

The timeline correlation operator $T$ defined in \eqref{eq:T-op} also immediately generalizes to $n$ timelines:
\[
T\left(\left|x\right\rangle \otimes\left|y\right\rangle \right)=\left|y\right\rangle \otimes\left|y\mnd x\right\rangle ,
\]
where now $\mnd$ represents subtraction modulo $n$. As in the 2-timeline case, the composite operator satisfies
\[
TU\left(\left|x\right\rangle \otimes\left|y\right\rangle \right)=\left|x\pld y\right\rangle \otimes\left|y\right\rangle .
\]
In particular, for the initial condition $y=1$ and $n\ge2$, it cyclically permutes the timeline labels as $i\mapsto i\pld1$ while leaving the external state unchanged. Thus all $n$ timelines form a single cycle, and no timeline is mapped to itself. Applying this operator, we find that the corresponding state at $t=0$ is
\[
T\left|\Psi\left(1\right)\right\rangle =TU\left|\Psi\left(0\right)\right\rangle =\sum^{n-1}_{i=0}\alpha_{i}\left|i\pld1\right\rangle \otimes\left|1\right\rangle .
\]
We thus get the consistency condition (allowing for an arbitrary overall phase)
\[
\sum^{n-1}_{i=0}\alpha_{i}\left|i\pld1\right\rangle \otimes\left|1\right\rangle =\e^{\ii\phi}\sum^{n-1}_{i=0}\alpha_{i}\left|i\right\rangle \otimes\left|1\right\rangle .
\]
Solving this as we did in Section~\ref{subsec:Correlation-and-consistency}, we find that the coefficients must satisfy
\[
\alpha_{i}=\e^{\ii\phi}\alpha_{i\pld1},\qquad\fa i\in\left\{ 0,\ldots,n-1\right\} .
\]
Using the cyclicity of indices, $\alpha_{i\pld n}=\alpha_{i}$, we get
\[
\alpha_{i}=\e^{n\ii\phi}\alpha_{i},\qquad\fa i\in\left\{ 0,\ldots,n-1\right\} ,
\]
which means that there are $n$ possible choices for $\phi$:
\[
\phi=\frac{2\pi k}{n},\qquad k\in\left\{ 0,\ldots,n-1\right\} .
\]
Since all coefficients have the same magnitude, the normalization condition \eqref{eq:n-normalization} implies they must all satisfy $\left|\alpha_{i}\right|=1/\sqrt{n}$. Note that this means $\left|\Psi\left(1\right)\right\rangle $ is maximally entangled, as in the 2-timeline case. Choosing without loss of generality $\alpha_{0}=1/\sqrt{n}$, we get the $n$ solutions
\[
\alpha_{0}=\frac{1}{\sqrt{n}},\qquad\alpha_{1}=\e^{-\ii\phi}\frac{1}{\sqrt{n}},\qquad\ldots,\qquad\alpha_{n-1}=\e^{-\left(n-1\right)\ii\phi}\frac{1}{\sqrt{n}},
\]
one solution for each choice of relative phase $\phi$. In other words, the consistent initial states of the time machine are the discrete \textbf{Fourier modes} of the timeline register
\[
\left|\chi_{k}\right\rangle \equiv\frac{1}{\sqrt{n}}\sum^{n-1}_{j=0}\e^{-2\pi\ii jk/n}\left|j\right\rangle ,\qquad k\in\left\{ 0,\ldots,n-1\right\} ,
\]
which are the eigenvectors of the cyclic timeline shift $i\mapsto i\pld1$ with eigenvalues $\e^{2\pi\ii k/n}$; the uniform superposition corresponds to the $k=0$ mode. Note that for $n=2$ we reproduce the solution of Section~\ref{subsec:Correlation-and-consistency}. As in the $n=2$ case, the relative phases have no effect on the results of this paper, so for simplicity, we will assume that all the coefficients are equal to $1/\sqrt{n}$. In that case, the states at $t=0$ and $t=1$ are given by
\begin{equation}
\left|\Psi\left(0\right)\right\rangle =\frac{1}{\sqrt{n}}\sum^{n-1}_{i=0}\left|i\right\rangle \otimes\left|1\right\rangle ,\qquad\left|\Psi\left(1\right)\right\rangle =\frac{1}{\sqrt{n}}\sum^{n-1}_{i=0}\left|i\right\rangle \otimes\left|i\pld1\right\rangle .\label{eq:n-timelines-t01}
\end{equation}
So far, the discussion has been completely abstract. We can make it more concrete using a variation of the macroscopic example of Section~\ref{subsec:A-macroscopic-example}, where the bombs turn out to be very ineffective, and Alice keeps trying more and more bombs in each timeline until she finally succeeds in killing her past self.

The state $\left|i\right\rangle $ of $\HH_{\mathrm{CTC}}$ corresponds to $i$ bombs inside the time machine ($\left|0\right\rangle $ means an empty time machine, as usual), and the state $\left|i\right\rangle $ of $\HH_{\mathrm{ex}}$ corresponds to Alice putting $i$ bombs inside the time machine. Alice starts in the state $\left|1\right\rangle $ because she initially (in all timelines) \textbf{intends} to use just one bomb. When she opens the door to the time machine, she discovers $i$ bombs inside.

If $i=0$, nothing happens and she continues to send 1 bomb to the past, as she intended.

If $0<i<n-1$, the $i$ bombs explode, but Alice survives. She then deduces that in the previous timeline, she put $i$ bombs inside the time machine, but this was not enough. Therefore, she decides to use $i+1$ bombs this time, in order to kill the Alice in the next timeline more effectively.

If $i=n-1$, the number of bombs is finally sufficient to kill Alice, and she dies; this means Alice's state changes to $\left|0\right\rangle $, since she is dead and therefore sends 0 bombs to the past. The reader can verify that this story is compatible with the abstract states and operators given above.

Note that in this example, as with the simpler 2-timeline examples in previous sections, the timelines are always \textbf{cyclic}: the last timeline (for example, the one where Alice is finally dead) always leads back to the first timeline (where the time machine is empty). Furthermore, we only considered a finite number of timelines. We leave the exploration of paradoxes with non-cyclic timelines and/or an infinite number of timelines to future work.

\section{\label{sec:Comparison-of-our}Comparison of Our Model to Deutsch's Model}

\subsection{Review of the D-CTC Model}

Deutsch's seminal 1991 paper \cite{Deutsch1991} presented a method for resolving time travel paradoxes within quantum mechanics, which has since become known as \textbf{the D-CTC model}, with the CTCs themselves referred to as Deutschian closed timelike curves or ``D-CTCs''.

There are some similarities between Deutsch's model and our entangled timelines model, which we will refer to as \textbf{the E-CTC model} or ``E-CTCs'' in this context. However, there are also significant differences, and we believe that the E-CTC model resolves some of the issues with the D-CTC model, as discussed below.

Deutsch considered 4 different paradoxes, formulated using qubits. Here we will focus specifically on Deutsch's ``paradox 1'', as it is (by design) mathematically equivalent to our own generic paradox model---except that the physical interpretation and resolution of the paradox are different.

In Deutsch's paradox, we consider a single qubit entering the interaction region, going into the time machine, emerging in the past, then interacting with its past self, and finally exiting the interaction region. The interaction between the young (past) and old (future) qubits is governed by the same unitary evolution operator $U$ we used in our model, as defined in \eqref{eq:unitary-U}:
\begin{equation}
U\left(\left|x\right\rangle \otimes\left|y\right\rangle \right)=\left|x\right\rangle \otimes\left|x\pld y\right\rangle ,\label{eq:D-int}
\end{equation}
where as before, $x,y\in\BBZ_{2}$ and $\pld$ denotes addition modulo 2. The state on the left of the tensor product represents the old (future) qubit, that is, the one that emerged from the time machine. The state on the right of the tensor product represents the young (past) qubit, that is, the one that entered from outside and has not yet traveled in time. Note that in Deutsch's original paper \cite{Deutsch1991}, the order of the systems in the tensor product is inverted, with the state of the time-traveling qubit on the right. We moved it to the left to conform with our notational conventions, where the timelines originate at the leftmost state in the tensor product and propagate gradually to the right (as described in Section~\ref{subsec:Spreading-of-the}).

The consistency paradox is now obtained by noticing that if the old qubit emerges from the time machine in the state $\left|x\right\rangle $, then after the interaction \eqref{eq:D-int}, the young qubit will enter the time machine in the state $\left|x\pld y\right\rangle $. If we assume that the state of the qubit does not change as it goes through the time machine, then the two states must be the same. This means that \textbf{classically}, we must have
\begin{equation}
x\pld y=x.\label{eq:D-clas}
\end{equation}
The only solution to this consistency condition is $y=0$, which means that the qubit enters the interaction region in the initial state $\left|0\right\rangle $. The value of $x$, on the other hand, is left unconstrained, so there are two self-consistent solutions. If the qubit emerges from the time machine in the state $\left|0\right\rangle $, it simply remains in the same state throughout. If it instead emerges in the state $\left|1\right\rangle $, it flips its younger self to $\left|1\right\rangle $, which then enters the time machine and closes the loop---a bootstrap paradox, as the flip exists only inside the loop, with no external origin. Such multiplicity of self-consistent solutions for the same initial data is a well-known feature of classical CTCs \cite{Friedman1990,Echeverria1991}. However, if we demand that the qubit starts in the initial state $\left|1\right\rangle $, corresponding to $y=1$, then the classical consistency condition \eqref{eq:D-clas} is not satisfied, and we have a consistency paradox.

The paradox is resolved in the quantum theory in the following way. Let $\left|\psi\right\rangle $ be the initial state of the young qubit as it enters the interaction region, and let $\rho$ be the density operator describing the state of the old qubit as it emerges from the time machine, which may be a mixed state. The initial joint state of the two qubits is then
\[
\rho\otimes\left|\psi\right\rangle \left\langle \psi\right|,
\]
and the state after the interaction is
\begin{equation}
U\left(\rho\otimes\left|\psi\right\rangle \left\langle \psi\right|\right)U^{\dagger}.\label{eq:D-after}
\end{equation}
To maintain consistency and resolve the paradox, Deutsch demands that the state of the qubit as it emerges from the time machine in the past is exactly the same as its state when it enters the time machine in the future. The former is, by definition, given by $\rho$, and the latter is obtained by tracing out the old qubit from the joint state \eqref{eq:D-after}. Therefore, the consistency condition is given by
\begin{equation}
\tr_{1}\left(U\left(\rho\otimes\left|\psi\right\rangle \left\langle \psi\right|\right)U^{\dagger}\right)=\rho,\label{eq:D-tr}
\end{equation}
where $\tr_{1}$ is the partial trace on the old qubit, so that only the state of the young qubit remains. We refer to this as \textbf{the D-CTC condition}.

Let us now take $\left|\psi\right\rangle =\left|1\right\rangle $, which is the initial state for which we get a paradox in the classical case. The unitary operator $U$ responsible for the interaction \eqref{eq:D-int} can be explicitly written in the outer product representation as
\begin{equation}
U=\sum_{x,y\in\BBZ_{2}}\left|x\right\rangle \otimes\left|x\pld y\right\rangle \left\langle x\right|\otimes\left\langle y\right|.\label{eq:D-gate}
\end{equation}
It is obvious that for computational-basis states,
\[
U\left(\left|x\right\rangle \left\langle x\right|\otimes\left|y\right\rangle \left\langle y\right|\right)U^{\dagger}=\left|x\right\rangle \left\langle x\right|\otimes\left|x\pld y\right\rangle \left\langle x\pld y\right|.
\]
Therefore, if we take
\begin{equation}
\rho=\hf\left(\left|0\right\rangle \left\langle 0\right|+\left|1\right\rangle \left\langle 1\right|\right),\label{eq:D-state}
\end{equation}
we will get after the interaction, given the initial condition $\left|\psi\right\rangle =\left|1\right\rangle $,
\begin{align*}
U\left(\rho\otimes\left|\psi\right\rangle \left\langle \psi\right|\right)U^{\dagger} & =\hf\left(U\left(\left|0\right\rangle \left\langle 0\right|\otimes\left|1\right\rangle \left\langle 1\right|\right)U^{\dagger}+U\left(\left|1\right\rangle \left\langle 1\right|\otimes\left|1\right\rangle \left\langle 1\right|\right)U^{\dagger}\right)\\
 & =\hf\left(\left|0\right\rangle \left\langle 0\right|\otimes\left|1\right\rangle \left\langle 1\right|+\left|1\right\rangle \left\langle 1\right|\otimes\left|0\right\rangle \left\langle 0\right|\right).
\end{align*}
Taking the partial trace with respect to the first qubit, we get
\[
\tr_{1}\left(U\left(\rho\otimes\left|\psi\right\rangle \left\langle \psi\right|\right)U^{\dagger}\right)=\hf\left(\left|1\right\rangle \left\langle 1\right|+\left|0\right\rangle \left\langle 0\right|\right)=\rho,
\]
and therefore the D-CTC condition \eqref{eq:D-tr} is satisfied. This means that the paradox is resolved if the time-traveling qubit is in a \textbf{maximally mixed state} of either $\left|0\right\rangle $ or $\left|1\right\rangle $ with probability 1/2 each. The qubit enters and exits the time machine in this state, and interacts with itself via the gate \eqref{eq:D-gate}, but this does not create any inconsistency, because it remains in the same mixed state throughout.

After the interaction, the young qubit goes into the time machine and therefore disappears. To find the output state, that is, the state of the old qubit as it leaves the interaction region, we need to trace out the young qubit:
\begin{equation}
\tr_{2}\left(U\left(\rho\otimes\left|\psi\right\rangle \left\langle \psi\right|\right)U^{\dagger}\right)=\hf\left(\left|0\right\rangle \left\langle 0\right|+\left|1\right\rangle \left\langle 1\right|\right).\label{eq:D-out}
\end{equation}
We see that the final state of the qubit is the same one that it had before the interaction, a mixed state of $\left|0\right\rangle $ and $\left|1\right\rangle $, with probability 1/2 each.

\subsection{\label{subsec:The-role-of}The Role of Parallel Timelines}

The mixed states \eqref{eq:D-state} and \eqref{eq:D-out} obtained in this paradox resolution can be interpreted as representing two different timelines. A single qubit approaches the time machine, and interacts with its older self. Then:
\begin{itemize}
\item In the first timeline, corresponding to the term $\left|0\right\rangle \left\langle 0\right|$ in \eqref{eq:D-state}, the young qubit is initially in the state $\left|1\right\rangle $ and the old qubit is in the state $\left|0\right\rangle $. They interact via \eqref{eq:D-int}, but neither changes its state in this interaction. The young qubit then goes into the time machine, still in the state $\left|1\right\rangle $, and disappears. The old qubit leaves the interaction region in the state $\left|0\right\rangle $, which corresponds to the term $\left|0\right\rangle \left\langle 0\right|$ in \eqref{eq:D-out}.
\item In the second timeline, corresponding to the term $\left|1\right\rangle \left\langle 1\right|$ in \eqref{eq:D-state}, the young qubit is again initially in the state $\left|1\right\rangle $, but the old qubit is now in the state $\left|1\right\rangle $; note that this qubit \textbf{came from the first timeline}. They interact via \eqref{eq:D-int}, and this time, the young qubit flips its state to $\left|0\right\rangle $. The young qubit then goes into the time machine in the state $\left|0\right\rangle $ and disappears; it, in fact, \textbf{emerges in the first timeline}. The old qubit leaves the interaction region in the state $\left|1\right\rangle $, which corresponds to the term $\left|1\right\rangle \left\langle 1\right|$ in \eqref{eq:D-out}.
\end{itemize}

This is perfectly consistent, but there is still a problem, because there is no timeline where time travel has not happened ``yet''; time travel happens in both timelines, just with qubits in different states. This makes it unclear how this paradox can be interpreted in terms of branching timelines, as there must be an ``original'' timeline for the ``new'' timeline to branch out of.

Deutsch himself was most likely aware of this issue, as when discussing the parallel timelines interpretation he chose to focus on a different paradox, ``paradox 3''. In this paradox, the qubits are interpreted as occupation numbers, indicating whether a particle is present or absent. Furthermore, another qubit is added to the mix, and the interaction is given by
\[
U\left(\left|x\right\rangle \otimes\left|y\right\rangle \otimes\left|y\pld1\right\rangle \right)=\left|x\right\rangle \otimes\left|x\pld y\right\rangle \otimes\left|x\pld y\pld1\right\rangle .
\]
Again, we remind the reader of our different notational convention, where the state of the time-traveling qubit is on the left, while in Deutsch's original paper it is on the right. Here, the occupation number on the left indicates whether or not a particle has traveled through the time machine, while the middle and right occupation numbers indicate whether a particle is traveling along one of two possible trajectories. Let us call the one in the middle \textbf{trajectory A} and the one on the right \textbf{trajectory B}. It is only trajectory B which actually goes into the time machine, which means that the classical consistency condition is
\[
x\pld y\pld1=x,
\]
therefore we must have $y=1$, while $x$ remains unconstrained. Choosing $x=0$ gives the initial state $\left|0\right\rangle \otimes\left|1\right\rangle \otimes\left|0\right\rangle $ where the particle is in trajectory A, which does \textbf{not} go into the time machine. Since no time travel occurs, there is also no paradox. Choosing $x=1$ gives a second self-consistent bootstrap solution. However, for $y=0$, corresponding to the initial state $\left|0\right\rangle \otimes\left|0\right\rangle \otimes\left|1\right\rangle $ where the particle is in trajectory B, which \textbf{does} go into the time machine, there is an inconsistency.

As with paradox 1, the inconsistency is resolved by demanding an appropriate modification of the D-CTC condition \eqref{eq:D-tr}:
\[
\tr_{1,2}\left(U\left(\rho\otimes\left|\psi\right\rangle \left\langle \psi\right|\right)U^{\dagger}\right)=\rho,
\]
where $\left|\psi\right\rangle $ is the joint state of trajectories A and B, and $\rho$ is the state of the time machine. Note that here $\tr_{1,2}$ traces out the time machine and trajectory A, so that only trajectory B remains, as it is the trajectory that goes into the time machine and emerges as $\rho$. The solution for the initial state $\left|\psi\right\rangle =\left|0\right\rangle \otimes\left|1\right\rangle $, corresponding to a particle in trajectory B, is the same as the solution \eqref{eq:D-state} for paradox 1:
\begin{equation}
\rho=\hf\left(\left|0\right\rangle \left\langle 0\right|+\left|1\right\rangle \left\langle 1\right|\right).\label{eq:D-state-3}
\end{equation}
Furthermore, the output state is given by
\begin{equation}
\tr_{3}\left(U\left(\rho\otimes\left|\psi\right\rangle \left\langle \psi\right|\right)U^{\dagger}\right)=\hf\left(\left|00\right\rangle \left\langle 00\right|+\left|11\right\rangle \left\langle 11\right|\right),\label{eq:D-out-3}
\end{equation}
where we traced out trajectory B, as it went into the time machine and disappeared.

Deutsch's paradox 3 is much more suitable for interpretation in terms of parallel timelines. A particle approaches the time machine in trajectory B, and then:
\begin{itemize}
\item In the first timeline, corresponding to the term $\left|0\right\rangle \left\langle 0\right|$ in \eqref{eq:D-state-3}, no particle emerges from the time machine. The incoming particle simply continues along trajectory B and enters the time machine. No particles are left, which corresponds to the term $\left|00\right\rangle \left\langle 00\right|$ in \eqref{eq:D-out-3}.
\item In the second timeline, corresponding to the term $\left|1\right\rangle \left\langle 1\right|$ in \eqref{eq:D-state-3}, a particle does emerge from the time machine; note that this particle \textbf{came from the first timeline}. The incoming particle interacts with this particle, which causes it to move to trajectory A. Therefore, it does not enter the time machine. Two particles are left, the incoming one and the one that came from the first timeline, which corresponds to the term $\left|11\right\rangle \left\langle 11\right|$ in \eqref{eq:D-out-3}.
\end{itemize}

The reason for adding an extra qubit is that Deutsch wanted to be able to describe something akin to the familiar Polchinski paradox \cite{Bishop2022}, where a billiard ball collides with its older self, and as a result, changes its trajectory from B to A, and fails to go into the time machine. A more technical reason is that properly describing both the input and output states using the same Hilbert space requires a total of 4 different states:
\begin{enumerate}
\item $\left|10\right\rangle $ corresponds to a particle in trajectory A,
\item $\left|01\right\rangle $ corresponds to a particle in trajectory B,
\item $\left|00\right\rangle $ corresponds to no particles in either trajectory,
\item $\left|11\right\rangle $ corresponds to a particle in each trajectory.
\end{enumerate}

In the generic paradox of the E-CTC model, as described in Section~\ref{subsec:The-generic-paradox}, we avoided this complication by using a formulation mathematically equivalent to Deutsch's paradox 1, except that instead of a single physical qubit interacting with itself, we interpret the qubit states as representing more abstract \textbf{logical} (true or false) states: whether the time machine is empty or not, and whether the external system will initiate time travel or not. In the E-CTC model, it is therefore perfectly fine to have essentially the same situation as in Deutsch's paradox 3, and yet only use 2 qubits to describe it.

\subsection{The Consistency Condition}

A large body of fascinating literature has been written about the D-CTC consistency condition \eqref{eq:D-tr}, studying its consequences and identifying potential issues \cite{Allen2014}. However, to our knowledge, the D-CTC condition itself has seldom been challenged.

Our main argument in this section is that the D-CTC condition \eqref{eq:D-tr} is a \textbf{sufficient} requirement for consistency, but not a \textbf{necessary} one. Imposing it is therefore a choice, and we argue that it is not the best choice if one is interested in the individual timelines, as it ends up partially obfuscating what is really going on. In the E-CTC model, we make a different choice: we require that the state at $t=1$ is related to the state at $t=0$ via the timeline correlation operator $T$ defined in \eqref{eq:T-op}.

This is a crucial difference between the D-CTC and E-CTC models, so let us emphasize it:
\begin{itemize}
\item Deutsch achieves consistency by demanding that the \textbf{reduced state} of the time-traveling particle is the same at $t=1$ and $t=0$, and ends up with a mixed state that is a statistical mixture of $\left|0\right\rangle $ and $\left|1\right\rangle $.
\item We instead obtain consistency by requiring that the \textbf{overall state} of the joint system at $t=1$, including not only the time machine itself but also the external system entangled with it, is suitably correlated with the overall state of the joint system at $t=0$, using the timeline correlation operator \eqref{eq:T-op}.
\end{itemize}

This condition allows us to describe not just the time-traveling qubit, but the \textbf{entire system} (and indeed, eventually, all systems in the causal future of the interaction), in terms of parallel timelines. Each decohered component of the superposition represents a different timeline, and as we have seen in Section~\ref{subsec:Additional-timelines}, the model can describe any finite number $n$ of possible timelines.

In the D-CTC model, we can use the evolution operator \eqref{eq:D-int} to find out how each individual timeline evolves. However, there is no well-defined way to find out how each timeline is connected to the next one. The mixed state \eqref{eq:D-state} does not contain all the information, so in our description of what is going on in each timeline at the beginning of Section~\ref{subsec:The-role-of}, we had to make an educated guess. This is trivial in the case of two timelines, but becomes more significant with additional timelines, as in Section~\ref{subsec:Additional-timelines}.

In the E-CTC model, we use an equivalent evolution operator \eqref{eq:unitary-U} to evolve each timeline, but in addition, by acting with the timeline correlation operator \eqref{eq:T-op} not on the overall state but on a single term in the superposition, we find how one timeline at $t=1$ is correlated with the state of the next timeline at $t=0$.

Therefore, the timeline correlation operator gives us a prescription to find the ``target'' timeline from the ``source'' timeline, which does not exist in Deutsch's model. This operator relates the joint state of the entangled system immediately before time travel to the corresponding separable state immediately after arrival in the past.

Furthermore, in the D-CTC model, it is unclear how the split into two timelines propagates further. In the E-CTC model, we have a clear picture of timelines propagating gradually through entanglement of additional systems in a well-defined way, as described in Section~\ref{subsec:Spreading-of-the}.

To understand the reason that the D-CTC condition is not necessary for consistency, consider again the states of the system at $t=0$ and $t=1$ in our generic model:

\[
\left|\Psi\left(0\right)\right\rangle =\frac{1}{\sqrt{2}}\left(\left|0\right\rangle +\left|1\right\rangle \right)\otimes\left|1\right\rangle ,
\]
\begin{equation}
\left|\Psi\left(1\right)\right\rangle =\frac{1}{\sqrt{2}}\left(\left|0\right\rangle \otimes\left|1\right\rangle +\left|1\right\rangle \otimes\left|0\right\rangle \right).\label{eq:entangled}
\end{equation}
We \textbf{could} require the D-CTC condition, in which case we would find that the state of $\HH_{\mathrm{CTC}}$ is the mixed state
\begin{equation}
\rho=\hf\left(\left|0\right\rangle \left\langle 0\right|+\left|1\right\rangle \left\langle 1\right|\right).\label{eq:mixed}
\end{equation}
This essentially tells us that the \textbf{reduced density operator} of the entangled state \eqref{eq:entangled}, after tracing out $\HH_{\mathrm{ex}}$, is given by the mixed state \eqref{eq:mixed}---which is perfectly correct, but conceals the fact that the state of the entire system is a pure entangled state.

When we take the partial trace of a pure composite state, the reduced state will be mixed if and only if the original state is entangled. Thus, in our paradox model, the maximally mixed D-CTC state is the reduced state of the pure entangled E-CTC state. But taking the reduced state of an entangled state \textbf{removes information} about the entanglement itself; it only tells us the probabilities for each of the possible states of one system, but cannot tell us anything about the corresponding correlated state of the other system, which we traced out.

As an example, consider the Bell states
\[
\left|\Phi^{+}\right\rangle =\frac{1}{\sqrt{2}}\left(\left|00\right\rangle +\left|11\right\rangle \right),\qquad\left|\Psi^{+}\right\rangle =\frac{1}{\sqrt{2}}\left(\left|01\right\rangle +\left|10\right\rangle \right).
\]
It is easy to see that the reduced density operator of the first qubit is \eqref{eq:mixed} in both cases. However, in the case of $\left|\Phi^{+}\right\rangle $, the second qubit will have the same value as the first qubit, while in the case of $\left|\Psi^{+}\right\rangle $, the second qubit will have the opposite value; the information about the correlation is thrown away completely when we trace out the second qubit, and cannot be restored.

Similarly, the D-CTC condition throws away the information about the entanglement of the time machine $\HH_{\mathrm{CTC}}$ with the external system $\HH_{\mathrm{ex}}$, which we believe is crucial for defining the timelines as separate ``worlds'' in the framework of the MWI. The E-CTC model instead imposes its consistency condition directly on the complete entangled state of the system, which contains the full information about the timelines, and therefore provides us with a more precise and well-defined picture of parallel timelines compared to the D-CTC model.

Another piece of information lost when taking the partial trace in the D-CTC model is the relative phase between the terms of the entangled state. Of course, this phase is expected to eventually become locally inaccessible due to decoherence \cite{Zurek2003b,Zurek2003a,Schlosshauer2007,Chua2025} in both models, but with one crucial difference.

In our paradox model, the D-CTC condition forces the time-traveling qubit to be in a maximally mixed state, which contains no information about the relative phase. On the other hand, in the E-CTC model, environmental decoherence results in the usual MWI picture where classical ``worlds'' emerge from purely unitary quantum mechanics, while the phase remains encoded in the global state.

Finally, let us also mention that in some of Deutsch's paradox examples, the solutions to the D-CTC condition are not unique. To select a solution uniquely, Deutsch conjectures the \textbf{maximum entropy rule}, which selects the solution with maximum entropy. In our 2-timeline model, the consistent solution is unique up to a relative minus sign, which does not affect the observables considered here (see Section~\ref{subsec:Correlation-and-consistency}). The $n$-timeline extension likewise has $n$ possible relative-phase patterns, so such a rule is not necessary in these examples. Moreover, when both pure and mixed solutions to the D-CTC condition are available, the maximum entropy rule will prefer a mixed solution, as pure states have zero entropy; in our model, the E-CTC states are always pure.

\subsection{Connections Between Timelines}

A different---perhaps more intuitive---way to explain our point in the previous section regarding loss of information about the correlations between $\HH_{\mathrm{CTC}}$ and $\HH_{\mathrm{ex}}$ is that in D-CTC it is impossible to know which timeline the time traveler will find themselves in when they step out of the time machine. In E-CTC, the timeline correlation operator gives us exactly that information; if the traveler is in the state $\left|y\right\rangle $ when they enter the time machine, they will exit the time machine in the timeline corresponding to the state $\left|y\right\rangle $:
\[
T\left(\left|x\right\rangle \otimes\left|y\right\rangle \right)=\left|y\right\rangle \otimes\left|y\mnd x\right\rangle .
\]
This becomes even more significant in the case of $n>2$ timelines, as described in Section~\ref{subsec:Additional-timelines}, where D-CTC would only tell us that the time machine is in the mixed state $\rho=I/n$ (where $I$ is the $n\times n$ identity matrix), while E-CTC tells us much more: that the pure states of the system at $t=0$ and $t=1$ are given by \eqref{eq:n-timelines-t01}
\[
\left|\Psi\left(0\right)\right\rangle =\frac{1}{\sqrt{n}}\sum^{n-1}_{i=0}\left|i\right\rangle \otimes\left|1\right\rangle ,\qquad\left|\Psi\left(1\right)\right\rangle =\frac{1}{\sqrt{n}}\sum^{n-1}_{i=0}\left|i\right\rangle \otimes\left|i\pld1\right\rangle ,
\]
where $\pld$ represents addition modulo $n$. After decoherence, we interpret each term in the entangled state at $t=1$ as a different timeline (equivalent to branches in the usual MWI formalism), and each timeline is enumerated by the state of the time machine, from $0$ to $n-1$. Crucially, the timeline correlation operator $T$ then tells us that if we enter the time machine at time $t=1$ in timeline $i$, which corresponds to the term $\left|i\right\rangle \otimes\left|i\pld1\right\rangle $ in the superposition of $\left|\Psi\left(1\right)\right\rangle $, we will find ourselves at time $t=0$ in timeline $i\pld1$, corresponding to the term $\left|i\pld1\right\rangle \otimes\left|1\right\rangle $ in the superposition of $\left|\Psi\left(0\right)\right\rangle $.

\subsection{Non-Linearity}

Both the D-CTC condition \eqref{eq:D-tr}
\[
\tr_{1}\left(U\left(\rho\otimes\left|\psi\right\rangle \left\langle \psi\right|\right)U^{\dagger}\right)=\rho
\]
and the E-CTC condition \eqref{eq:consistency-condition}
\[
TU\left|\Psi\left(0\right)\right\rangle =\e^{\ii\phi}\left|\Psi\left(0\right)\right\rangle
\]
are fixed-point conditions. However, they induce different types of evolution. In the D-CTC model, the solution $\rho$ depends on the initial state $\left|\psi\right\rangle $ of the external system; while the condition itself is linear in $\rho$, the induced evolution of the external system is \textbf{non-linear} in its initial state. It is well-known that this D-CTC non-linearity leads to many surprising consequences, including:
\begin{itemize}
\item Non-orthogonal states can be distinguished perfectly \cite{Brun2009}, violating the Holevo bound \cite[chapter 11]{Wilde2017}.
\item Arbitrary states can be cloned \cite{Ahn2013,Brun2013}, violating the no-cloning theorem \cite{Wootters1982}.
\item Classical computers can be as powerful as quantum computers \cite{Aaronson2009}.
\item The evolution of the input state is discontinuous \cite{DeJonghe2010}.
\item Superluminal signaling is possible \cite{Bub2014}.
\end{itemize}

The E-CTC condition, on the other hand, is just an eigenvalue equation. Since the operator $TU$ acts as a permutation of the basis states of the time machine for each basis state of the external system, the uniform superposition
\[
\left|\chi_{0}\right\rangle \equiv\frac{1}{\sqrt{n}}\sum^{n-1}_{i=0}\left|i\right\rangle
\]
of time machine states satisfies the consistency condition \eqref{eq:consistency-condition}, with $\phi=0$, for \textbf{any} initial state $\left|\psi\right\rangle $ of the external system. The initial state of the time machine can therefore be chosen independently of the input, and the evolution of the composite system, given by acting with $U$ on $\left|\chi_{0}\right\rangle \otimes\left|\psi\right\rangle $, is manifestly \textbf{linear} in $\left|\psi\right\rangle $. In contrast, in the D-CTC model with the same interaction, the solution of the consistency condition \eqref{eq:D-tr} depends on the input state: for $\left|\psi\right\rangle =\left|1\right\rangle $ the solution is the maximally mixed state \eqref{eq:D-state}, while for $\left|\psi\right\rangle =\left(\left|0\right\rangle +\left|1\right\rangle \right)/\sqrt{2}$ the solution is the pure state $\left|\psi\right\rangle \left\langle \psi\right|$ itself.

Moreover, this argument is not limited to the specific interaction $U$ considered in this paper. Recall from Section~\ref{subsec:Correlation-and-consistency} that the timeline correlation operator $T$ was chosen precisely such that the composite operator $TU$ permutes the timelines, and thus defines the connections between them: each timeline leads to exactly one timeline. Denoting this permutation operator by $W$, defined by
\[
W\left(\left|x\right\rangle \otimes\left|y\right\rangle \right)\equiv\left|x\pld y\right\rangle \otimes\left|y\right\rangle ,
\]
we see that for \textbf{any} interaction $U$, the corresponding timeline correlation operator is given by $T=WU^{\dagger}$. Since the consistency condition \eqref{eq:consistency-condition} only involves the composite operator $TU=W$, the uniform superposition of time machine states remains a consistent initial state for any initial state of the external system and any choice of $U$.

For each fixed interaction $U$, the induced evolution
\[
\left|\psi\right\rangle \mapsto U\left(\left|\chi_{0}\right\rangle \otimes\left|\psi\right\rangle \right)
\]
is a linear isometry independent of the input state $\left|\psi\right\rangle $. Therefore, all the standard results of quantum mechanics still apply: inner products are preserved, so non-orthogonal states cannot be distinguished; the proofs of the no-cloning and no-signaling theorems remain valid; the evolution is continuous in the input state; and no computational power is gained beyond standard quantum mechanics, as the time machine merely acts as an ancillary system prepared in a fixed state. We conclude that none of the effects listed above can occur in the E-CTC model. We leave a more in-depth exploration of this argument to future work.

\subsection{The Output State}

Similar considerations arise with regard to the output state of the system. Recall that in the D-CTC model, the output state of the two qubits is given by the mixed state \eqref{eq:D-out}:
\begin{equation}
\tr_{2}\left(U\left(\rho\otimes\left|\psi\right\rangle \left\langle \psi\right|\right)U^{\dagger}\right)=\hf\left(\left|0\right\rangle \left\langle 0\right|+\left|1\right\rangle \left\langle 1\right|\right).\label{eq:D-out2}
\end{equation}
The idea here is that we are tracing out qubit 2 because it went into the time machine and ``disappeared''. In the E-CTC model, we instead consider the pure entangled state \eqref{eq:entangled},
\begin{equation}
\left|\Psi\left(1\right)\right\rangle =\frac{1}{\sqrt{2}}\left(\left|0\right\rangle \otimes\left|1\right\rangle +\left|1\right\rangle \otimes\left|0\right\rangle \right),\label{eq:entangled2}
\end{equation}
as the ``output state'', \textbf{without tracing anything out}.

More importantly, tracing out part of the entangled state means we are left with a \textbf{reduced state}, which destroys information about the entanglement and therefore about the timelines. If we wish to determine the output state of the system, it makes more sense to look at the \textbf{entire state} \eqref{eq:entangled2}, where we can clearly see the states of both the time machine and the external system in each individual timeline and the correlation between them.

To illustrate this in different terms, let us recall from Section~\ref{subsec:Quantum-mechanics-without} that according to the MWI, if Alice measures a qubit, then the composite state of Alice and the qubit will be given by \eqref{eq:MWI-branching}:
\begin{equation}
\left|\Psi\left(1\right)\right\rangle =a\left|0\right\rangle \otimes\left|\textrm{Alice measured 0}\right\rangle +b\left|1\right\rangle \otimes\left|\textrm{Alice measured 1}\right\rangle .\label{eq:MWI-branching2}
\end{equation}
One might argue that the output state of the qubit is given by tracing out Alice's state; that is,
\[
\rho=\left|a\right|^{2}\left|0\right\rangle \left\langle 0\right|+\left|b\right|^{2}\left|1\right\rangle \left\langle 1\right|.
\]
However, this, again, discards information about the entanglement, and therefore about the different ``worlds''. A more sensible way to describe the output state of the qubit is to use the entire expression \eqref{eq:MWI-branching2}, which tells us that a branching has occurred such that in one branch the qubit's state is $\left|0\right\rangle $ and Alice's state is $\left|\textrm{Alice measured 0}\right\rangle $, while in the other branch the qubit's state is $\left|1\right\rangle $ and Alice's state is $\left|\textrm{Alice measured 1}\right\rangle $. It is meaningless to ask what ``the state of the qubit'' is in the first place, because there is more than one answer to this question---it has \textbf{a different state in each branch}.

In exactly the same way, the mixed state \eqref{eq:D-out2} we obtained from the D-CTC model contains no information about the entanglement, so we know what the possible states are, but not which timeline each state belongs to. It is therefore better to say that the output state is given by \eqref{eq:entangled2}, which provides complete information about the state of each component of the system in each timeline.

\subsection{The Relative States}

In the MWI formalism, the state of each subsystem in each individual branch is called the \textbf{relative state} \cite{Everett1957,Everett2012}. We can adapt this to our model as follows. The composite state after the external system interacts with the time machine is given by \eqref{eq:entangled2}:
\[
\left|\Psi\left(1\right)\right\rangle =\frac{1}{\sqrt{2}}\left(\left|0\right\rangle \otimes\left|1\right\rangle +\left|1\right\rangle \otimes\left|0\right\rangle \right).
\]
In E-CTC, the state on the left of the tensor product ($\HH_{\mathrm{CTC}}$) not only describes the state of the time machine, but also enumerates each timeline. So we can immediately find the relative state of the external system ($\HH_{\mathrm{ex}}$) in each timeline by inspection, as it is the state on the right of the tensor product in the term corresponding to that timeline:
\begin{equation}
|\Psi^{\left(0\right)}_{\mathrm{ex}}\rangle=\left|1\right\rangle ,\qquad|\Psi^{\left(1\right)}_{\mathrm{ex}}\rangle=\left|0\right\rangle .\label{eq:branch-states}
\end{equation}
For a more rigorous derivation, we can define the partial inner product map
\[
R_{i}:\HH_{\mathrm{CTC}}\otimes\HH_{\mathrm{ex}}\to\HH_{\mathrm{ex}},\qquad R_{i}\left(\left|x\right\rangle \otimes\left|y\right\rangle \right)\equiv\langle i|x\rangle|y\rangle,
\]
where $i\in\left\{ 0,1\right\} $ is the timeline index. Using this map, we can define the relative state in timeline $i$ as
\begin{equation}
|\Psi^{\left(i\right)}_{\mathrm{ex}}\rangle\equiv\frac{R_{i}\left|\Psi\left(1\right)\right\rangle }{\left\Vert R_{i}\left|\Psi\left(1\right)\right\rangle \right\Vert },\label{eq:relative-state}
\end{equation}
which then yields \eqref{eq:branch-states}. More generally, for the case of $n$ timelines as described in Section~\ref{subsec:Additional-timelines}, we have from \eqref{eq:n-timelines-t01}:
\[
\left|\Psi\left(1\right)\right\rangle =\frac{1}{\sqrt{n}}\sum^{n-1}_{i=0}\left|i\right\rangle \otimes\left|i\pld1\right\rangle .
\]
The relative states will therefore be, for each timeline $i$:
\[
|\Psi^{\left(i\right)}_{\mathrm{ex}}\rangle=\left|i\pld1\right\rangle .
\]
We see that in E-CTC, the external system is in a well-defined \textbf{pure} state in each timeline, which can be found using \eqref{eq:relative-state}. We are able to find this state because in our model we know the full state of the composite system at $t=1$; in D-CTC the most we can do is find the (mixed) state of the external system, which does not tell us which of the states in the ensemble corresponds to which timeline.

\subsection{Summary: D-CTC vs. E-CTC}

Deutsch's original D-CTC model introduced the intriguing concept of resolving time travel paradoxes by using the Everett interpretation to facilitate parallel timelines (referred to by Deutsch as multiple ``universes'', see \cite{Deutsch1991} (pages 3206--3207) and \cite{Deutsch1994}). The novelty of the E-CTC model is in providing an alternative consistency condition which acts directly on entangled pure states, rather than on reduced mixed states. This creates a clearer practical definition of the parallel timelines themselves:
\begin{itemize}
\item Each timeline corresponds to a different decohered component of the entangled state.
\item Each timeline evolves individually via the evolution operator $U$ defined in \eqref{eq:unitary-U}.
\item Each timeline correlates with the next one via the timeline correlation operator $T$ defined in \eqref{eq:T-op}.
\item Consistency is guaranteed by demanding that the entire entangled state at $t=1$ is related to the state at $t=0$ via the timeline correlation operator $T$, or equivalently, that the state at $t=0$ is related to itself by the composite operator $TU$.
\item The timelines gradually spread out to more and more systems by entanglement, as shown in Section~\ref{subsec:Spreading-of-the}.
\end{itemize}

These aspects of the timelines are obscured in the D-CTC model because it is defined in terms of reduced states obtained via a partial trace, which destroys information about the entanglement, and therefore about the timelines themselves. Furthermore, as our model retains the linear evolution of standard quantum mechanics, it can avoid the surprising consequences of the non-linear evolution in Deutsch's original model, as detailed above.

Further analysis of the D-CTC model in the context of the MWI can be found in Ref.~\cite{Wallace2012} (Section 10.6 and references therein). We also refer the reader to Ref.~\cite{Tolksdorf2021}, where it is shown that the D-CTC model can be reproduced in classical statistical systems due to the use of mixed states, and to Ref.~\cite{Dunlap2015}, which discusses whether the mixed states in the D-CTC model can properly be interpreted as separate branches in the MWI. These issues do not arise in the E-CTC model, as it is defined in terms of pure states.

Finally, it is important to note that, at the level of the underlying 2-qubit system and interaction $U$, our generic paradox is mathematically identical to Deutsch's paradox 1. The models differ in their consistency conditions, but other than that, we do not make any physical assumptions that Deutsch does not make. We interpret the qubit states differently in order to tell what we believe to be a more compelling and intuitive ``story''; instead of talking about a quantum computational network, we talk about a ``time machine'' and an ``external system''. However, both of these systems are simply abstract, logical qubits; the first qubit is an occupation number which tells us whether the time machine is empty or not, and the second is a control bit that tells us whether the external system will initiate time travel or not. The underlying description is still that of a 2-qubit system with an implied, abstract CTC modeled as a ``negative delay component'', similarly to Deutsch's original model.

\section{Conclusions}

In this paper, we introduced a new model for the resolution of time travel paradoxes using parallel timelines, called the entangled timelines or E-CTC model. We showed that this model resolves generic time travel paradoxes formulated using an abstract 2-qubit system, which can be mapped onto more complicated microscopic or macroscopic systems.

The usual picture of parallel timelines presented in the literature is that of a universe splitting into two physical copies of itself. However, in our model, the timelines are not physically distinct universes but rather emergent structures which exist due to entanglement between systems and the subsequent decoherence, similarly to the ``worlds'' or ``branches'' of the Everett or many-worlds interpretation. Timelines are local structures within a single universe, propagating gradually through the causal future of the interaction as more and more systems become entangled.

In contrast with Deutsch's familiar D-CTC model, our model only involves pure states and defines timelines explicitly as emerging from entanglement and decoherence. We argue that the D-CTC condition, which requires that the reduced states of the time-traveling qubit are the same before and after time travel occurs, is sufficient but not necessary for consistency. We interpret the D-CTC mixed states resulting from this condition as reduced states of the E-CTC entangled states; this reduction discards information about the entanglement and therefore obscures the parallel timelines emerging from that entanglement.

The entangled timelines model is formulated using two operators: a unitary evolution operator $U$, which entangles the state of the time machine with the state of the external system and produces components that become timelines once decoherence suppresses interference between them, and a timeline correlation operator $T$, which correlates the joint state of the two systems immediately before time travel with the corresponding joint state immediately after arrival in the past.

Consistency is obtained in our model by acting on the initial joint superposition state with $U$ to evolve it from $t=0$ to $t=1$, then acting with $T$ to obtain the corresponding state at $t=0$, and finally requiring that the latter be equal to the initial state up to an overall phase. This condition is sufficient to maintain consistency and resolve the paradox.

In addition, the evolution within each timeline from $t=0$ to $t=1$ is defined by the action of $U$ on a particular term in the superposition, and the correlation from one timeline at $t=1$ to the next timeline at $t=0$ as a result of time travel is given by the action of $T$ on that same term.

Our model can describe any finite number of timelines, as long as they are cyclic; that is, the final timeline is linked back to the first one via the correlation operator $T$. Additionally, our model is mostly conceptual, with qubit states serving as placeholders for real physical states.

Possible directions for future work include:
\begin{itemize}
\item Developing a more general mathematical formulation of the model and characterizing its consistent solutions for general $U$ and $T$;
\item Investigating extensions involving infinitely many timelines or non-cyclic timeline structures;
\item Studying what happens when a subsystem entangled with another system enters the time machine, or when more than one time machine is present;
\item Formulating the model in continuous time, using quantum fields on a concrete spacetime containing CTCs instead of abstract qubits;
\item Studying the decoherence process in more detail, including the case of a finite environment, where decoherence is incomplete;
\item Determining whether the model can be accommodated by interpretations of quantum mechanics other than the MWI;
\item Exploring the relation of the model to other approaches to quantum causality, such as the process matrix formalism \cite{Oreshkov2012,Araujo2017};
\item Investigating the analogy between the consistency condition of the model and the resonance condition of a cavity, motivated by models which describe time travel as a feedback loop \cite{Greenberger2005,Czachor2019};
\item Comparing the operational and information-theoretic consequences of E-CTC and D-CTC consistency;
\item Determining under what conditions D-CTC states can be obtained as reduced states of E-CTC states.
\end{itemize}

\section{Acknowledgments}

Barak Shoshany would like to thank Amos Ori for helpful and stimulating discussions. This research was supported by the Mitacs Globalink Research Internship and by the Natural Sciences and Engineering Research Council of Canada (NSERC), RGPIN-2024-04063.

\printbibliography
\end{document}